\documentclass[11pt, a4paper, twocolumn]{article}
\usepackage{xurl}
\usepackage[super,comma,sort&compress]{natbib}
\usepackage{abstract}

\usepackage{lipsum}
\usepackage{amsmath}
\usepackage{braket}
\usepackage{multirow}
\usepackage[hidelinks]{hyperref}
\usepackage{xcolor}
\usepackage{graphicx}
\usepackage{authblk}
\usepackage{caption}
\usepackage{subcaption}
\usepackage[
left=1.00cm,
right=1.00cm,
top=2.00cm,
bottom=2.00cm
]{geometry}

\date{}

\title{Evolutionary computation for adaptive quantum device design}

\author[1]{Luke Mortimer}
\author[2]{Marta P. Estarellas}
\author[1]{Timothy P. Spiller}
\author[1]{Irene D'Amico}

\affil[1]{Department of Physics, University of York, York, YO10 5DD, United Kingdom}
\affil[2]{National Institute of Informatics, 2-1-2 Hitotsubashi, Chiyoda-ku, Tokyo 101-8430, Japan}

\newcommand{\abstractText}{\noindent
As Noisy Intermediate-Scale Quantum (NISQ) devices grow in number of qubits, determining good or even adequate parameter configurations for a given application, or for device calibration, becomes a cumbersome task. An evolutionary algorithm is presented here which allows for the automatic tuning of the parameters of any arrangement of coupled qubits, to perform a given task with high fidelity. The algorithm's use is exemplified with the generation of schemes for the distribution of quantum states and the design of multi-qubit gates. The algorithm is demonstrated to converge very rapidly, yielding unforeseeable designs of quantum devices that perform their required tasks with excellent fidelities. Given these promising results, practical scalability and application versatility, the approach has the potential to become a powerful technique to aid the design and calibration of NISQ devices.
}

\usepackage{hyperref}
\hypersetup{colorlinks=true, urlcolor=blue, linkcolor=blue, citecolor=blue}

\begin{document}


\twocolumn[
  \begin{@twocolumnfalse}
    \maketitle
    \begin{abstract}
      \abstractText
      \newline
      \newline
    \end{abstract}
  \end{@twocolumnfalse}
]


\section{Introduction}\label{sec:introduction}

Quantum technologies have already extensively manifested their potential to impact a wide spectrum of fields. The first proof-of-principle demonstration of quantum advantage in terms of computational power, also known as \emph{quantum supremacy}, \cite{preskill2012quantum} has already been achieved, \cite{Arute2019} satellite-based quantum-protected key sharing (quantum key distribution, QKD) has been realized, \cite{Liao2017} and early prototypes of a first quantum internet are starting to emerge. \cite{Wehnereaam9288} As the qubit-number and complexity of quantum technology devices increases, so does the number of their relevant parameters and, correspondingly, the size of the parameter space to investigate. Consequently, calibrating the devices and/or determining parameter values suitable to perform a desired task is becoming a very complex challenge.

At the same time, the application range is so varied that there is no established preferred physical hardware for such early devices and hybridized approaches towards technologies are currently being undertaken. \cite{Kurizki_2015} With quantum devices being so diverse and heterogeneous, spin networks form a very convenient mathematical model, able to capture the quantum dynamics of any arrangement of two-level quantum systems coupled to each other, independent of the actual physical implementation. \cite{nikoBook} The specificity of each physical system is instead captured by the network topology, couplings and the energy scale of the parameters. Because of this, the simulation of quantum chips under the spin network formalism has proven to be a useful test-bed for the study and design of new quantum hardware and its applications. 

Spin networks have been engineered to allow for quantum state transfer, \cite{christandl2004,niko2004,wang2011all,karbach2005,vinet2011} to present topologically protected states, \cite{estarellas2016, Andrea2016, Longhi2019} or to act as quantum gates, \cite{damico2007_2, spiller2007, estarellas2017robust, Apollaro2019, ronke2011knitting, tserkovnyak2011universal, landahl2004information} amongst a spectrum of applications. However, to achieve a desired dynamical behaviour the ability to choose a suitable network topology is required, as well as calibration of the system parameters, such as the interaction energies, or \emph{couplings}, between the different nodes (qubits) of the network. Whilst available technology already allows for excellent control of such parameters in the laboratory, \cite{Arute2019} determining suitable tuning, as well as an appropriate network topology, in the design of quantum devices is a highly non-trivial task for complex systems beyond toy-models. The number of possibilities and combinations is so vast that determining good or even adequate solutions for a given task can be not only cumbersome, but also counter-intuitive. 

To circumvent this, for the design of quantum devices via the engineering of spin networks, we here propose the use of tools common in evolutionary computation. \cite{Eiben2015} This computational paradigm was originally proposed in the 1970s with the idea of using concepts of natural evolution to solve hard computational tasks in optimization, design and modelling. \cite{Eiben2003} Since then, it has proven to be a powerful tool for problems that do not necessarily require optimal solutions, but instead can utilize appropriate approximations to these. To this purpose, genetic algorithms are one of the most popular techniques and have widely been applied to solve large engineering problems, ranging from antenna design, \cite{hornby2006automated} and complex aerodynamic modelling \cite{evans2017aerodynamic} to the improvement of artificial neural networks \cite{Yao1999} and the automatic identification of analytical equations underlying physics phenomena. \cite{Schmidt2009} The design of such systems is characterized by the number and complexity of their degrees of freedom, something that makes the number of possible configurations exponentially large. Genetic algorithms are highly parallelizable meta-heuristic optimization techniques able to efficiently cover such large search spaces and thus find approximate solutions. \cite{whitley1994genetic} Quantum systems presents similarities to these problems in terms of complexity, making genetic methods promising candidates to automate the search for appropriate design solutions. To date, such methods have been scarcely used in this field, with only one example to our knowledge. \cite{dominguez2015quantum}

Here we design a genetic algorithm capable of identifying optimal tuning parameters of a spin network to achieve any given quantum information task. Our algorithm is general and could be applied to any given problem of this sort. To exemplify its use, here we focus on two tasks: the engineering of quantum devices for quantum state distribution and the design of multi-qubit gates.

We not only demonstrate that the proposed automated technique may find new system configurations that were previously unknown, but we also show that it can do this very rapidly. Machine learning algorithms have recently proven useful for the tuning of semiconductor quantum devices \cite{moon2020,esbroeck2020} with a runtime of approximately 70 minutes for the tuning of 8 experimental parameters (gate voltages of a double quantum dot). Our investigations show our approach based on genetic algorithms to be a promising alternative for the tuning and identification of unforeseeable designs of quantum devices, with examples of runtimes as short as 5 seconds to optimize 10 model parameters up to a fidelity of 99.7\%.

\section{Spin network model}\label{sub:spinchain}

We consider a general spin network of $N$ sites (also referred to as spins, nodes or qubits) that can be described by the following time-independent XXZ-Heisenberg Hamiltonian:

\begin{equation}
\begin{split}
    \hat{H} = \sum_{i < j} &J_{ij}\left(\ket{1}\bra{0}_i \otimes \ket{0}\bra{1}_j + h.c.\right) \\
    &+ \sum_{i=1}^N \epsilon_i \ket{1}\bra{1}_i \\
    &+ \sum_{i < j} \alpha J_{ij}\left(\ket{1}\bra{1}_i \otimes \ket{1}\bra{1}_j\right)
    \label{eqn:hamil}
\end{split}
\end{equation}

with $J_{ij}$ being the real-valued coupling between sites $i$ and $j$, and $\epsilon_i$ the on-site energies. In our encoding, we consider the injection of an excitation to be the creation of a spin ``up'', $\vert 1\rangle$, in a system that has initially been prepared to have all the spins ``down'', $\vert 0\rangle$. The interaction term is proportional to the constant dimensionless scaling factor $\alpha$. For $\alpha\neq 0$, this term represents the interaction energy between two excitations \cite{ronke2011effect} and thus affects the dynamics of subspaces containing at least two excitations. The topology of the network is defined by the non-zero elements of $J_{ij}$. The values of $J_{ij}$ and of $\epsilon_i$ are the targets of the optimisation. Throughout this paper, $J_\text{max}$ indicates the maximum value of the couplings for a given system.

Once the parameters are set, we obtain the eigenvectors and eigenvalues of the Hamiltonian matrix by direct diagonalisation. Then any chosen initial state is decomposed into the eigenvectors, which are each evolved via the unitary operator $U=e^{-\frac{i}{\hbar}E_it}$, with $E_i$ being the corresponding eigenvalue. This allows us to obtain the overall state of the system at any time without loss of accuracy.

We use the fidelity as a measure to assess how close the state of the system at a given time $|\Psi(t)\rangle$ is to the specific target state $|\Psi_\text{target}\rangle$ required for our task. The fidelity $F(t)$ ranges between zero and unity, with maximum fidelity representing a perfect overlap between target and actual state:

\begin{equation}
    \text{F}(t)=|\braket{\Psi_\text{target}|\Psi(t)}|^2.
    \label{eqn:fidelity}
\end{equation}

\section{Genetic algorithm}\label{sub:method}

\begin{figure*}
    \centering
    \includegraphics[scale=0.85]{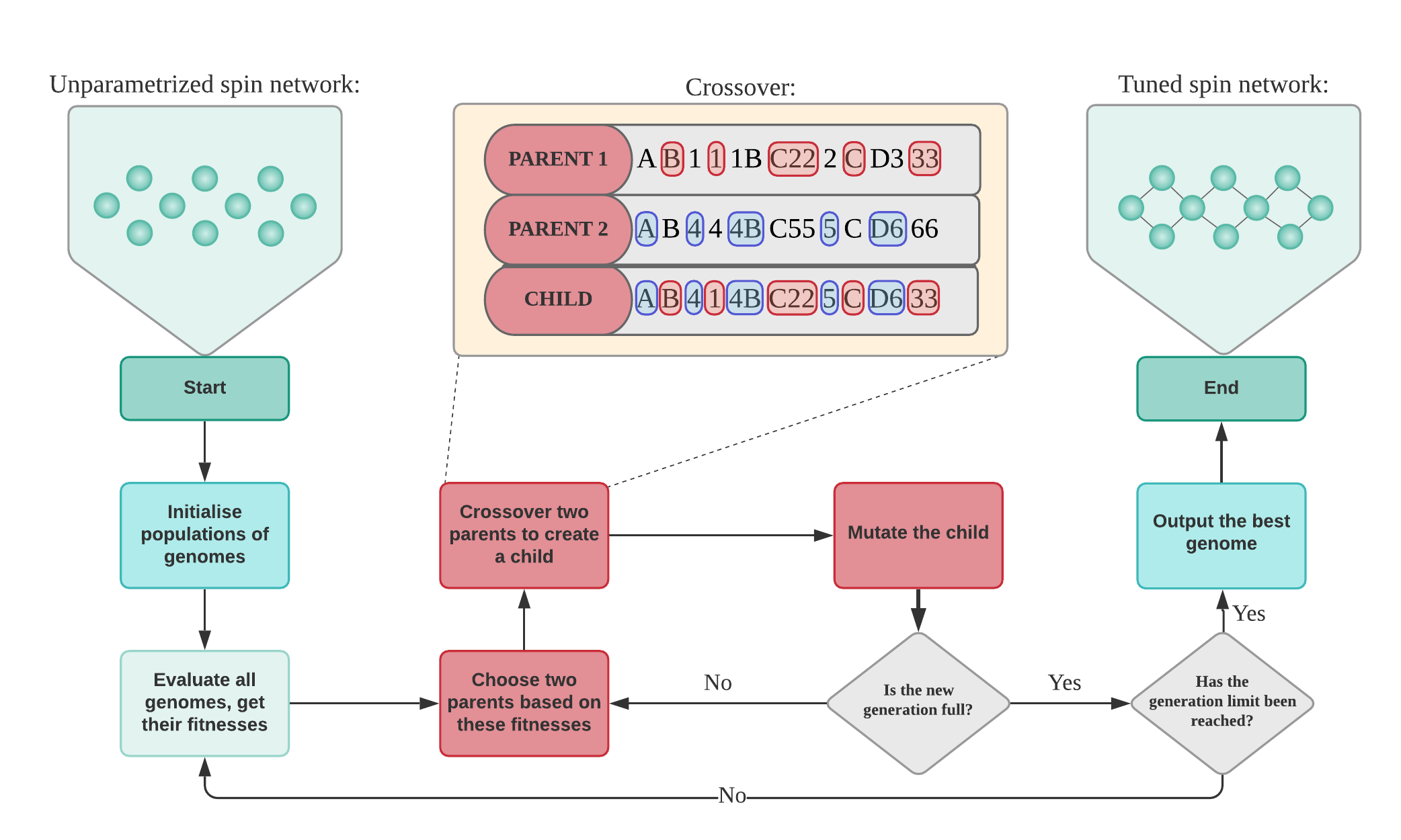}
    \caption{Schematic representation of our genetic algorithm. In the central panel we present a diagram displaying the method used to crossover two genomes. For each character in the child genome, the corresponding character from one of the parents is used, with each having equal probability.}
    \label{fig:flow}
\end{figure*}

General genetic algorithms rely on the evaluation of different parametrizations of a system's degrees of freedom to perform a given task. This evaluation is done through the calculation of a ``fitness'' score, which indicates how close a given parametrization is to optimality. In our implementation, the parameters to be tuned are the set of non-zero coupling energies of Equation \ref{eqn:hamil}.

The flowchart in Figure \ref{fig:flow} shows a schematic of the route followed by our genetic algorithm. The first and most critical step to design a genetic algorithm is to define a proper structure of what is called the ``genome''. As in DNA, the genome is represented as a string containing the mutable information of a system; this is, its degrees of freedom. The algorithm starts with a set of different genomes being evaluated according to a fitness function. After this evaluation, the better genomes are favoured to combine (``crossover'') with other successful genomes, based on their fitness scores, to form the next generation. After crossing-over two genomes, a random modification is then made to the resulting genome to allow for new and unique solutions to be found. \cite{thede2004introduction}

\subsection{A Genome For a Spin Network}

In order to properly utilize the features of a genetic algorithm, it is important to find an effective representation of the spin network in a standardized notation, so that it can be easily passed as an argument to the various functions of the program. As such, it should contain all of the system's relevant information, whilst also being easy to store, modify and transfer. The genome we chose to use here is represented as a fixed-size linear string of standard ASCII characters, split into various sections representing information useful for the different functions. There also exist some optional features that can be added into the string to specify more unique requirements.

In our genome, letters are used to represent the sites and their corresponding single-excitation basis vectors, whilst couplings energies are represented as integers. For example, $AB500$ would represent a coupling strength of 500 between sites A and B of a spin network, relative to any other specified couplings. A bra-ket $\braket{...|...}$ at the start specifies the initial and target states of the system's protocol, which is the information necessary to evaluate the performance of such a genome through its fidelity (Equation \ref{eqn:fidelity}). In this bra-ket, two letters placed adjacent are treated as the tensor product of the two single excitation basis vectors, for instance  $\bra{AB}=\bra{11}_{AB}$, thus allowing the use of multiple excitation subspaces. Superpositions can be described through addition, with an optional phase ($-1$, $i$, $(1+2i)$, etc.), such that $\ket{A+iB} = \frac{1}{\sqrt{2}}\left(\ket{10}_{AB}+i\ket{01}_{AB}\right)$. Note that in the algorithm all state vectors are automatically normalized, so normalization factors are left out from the genome.

\begin{figure}
    \centering
    \includegraphics[width=0.95\linewidth]{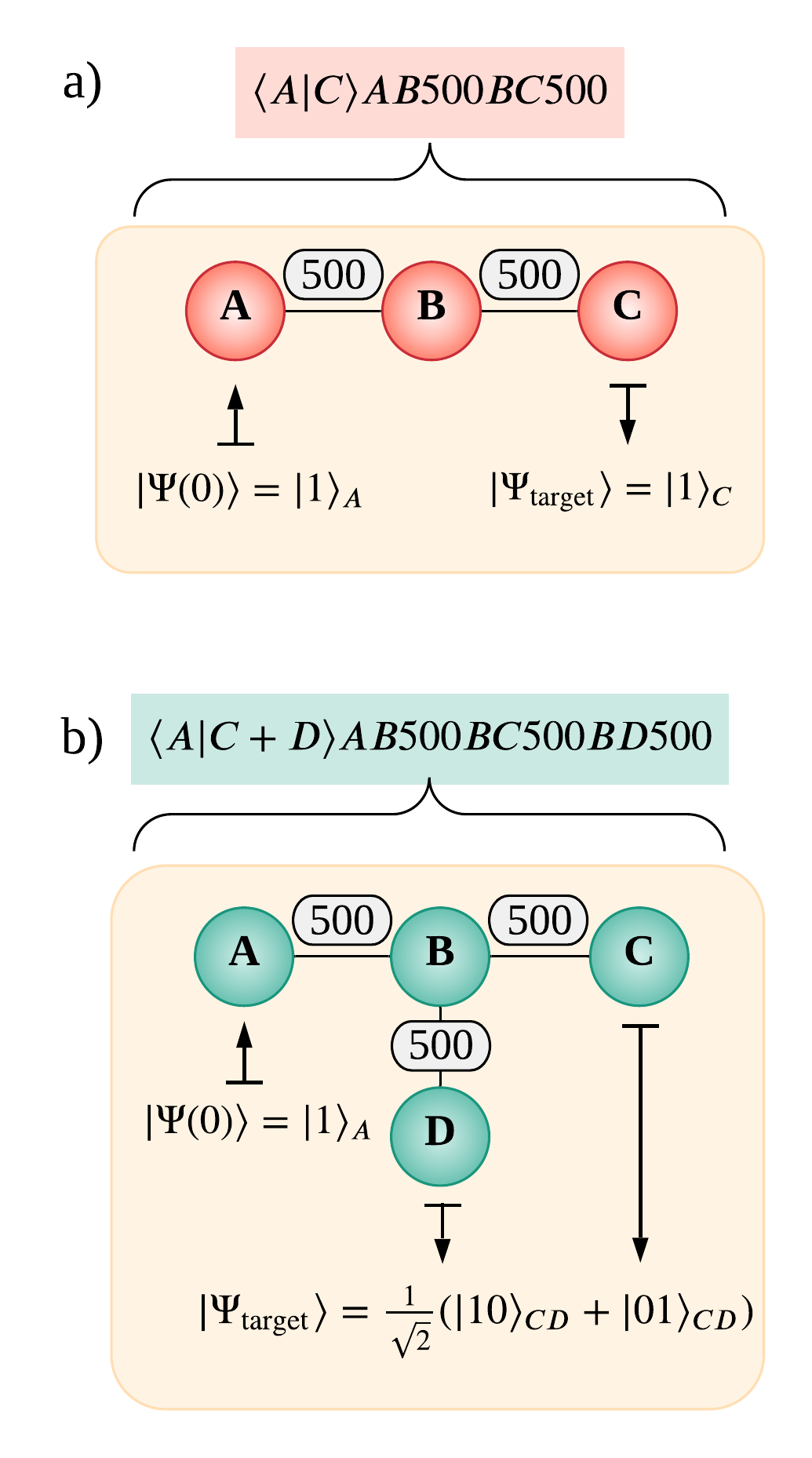}
    \caption{Examples of the construction of a genome in our algorithm. \textbf{a)} Uniformly coupled ($J=500$) $N$=3 spin chain with initial state $\vert\Psi(0)\rangle=\vert100\rangle_{ABC}$ and target state $\vert\Psi_{\text{target}}\rangle=\vert001\rangle_{ABC}$. \textbf{b)} Uniformly coupled ($J=500$) spin network with initial state $\vert\Psi(0)\rangle=\vert1000\rangle_{ABCD}$ and target state $\vert\Psi_{\text{target}}\rangle=\ket{00}_{AB}\otimes\frac{1}{\sqrt{2}}(\ket{10}_{CD}+\ket{01}_{CD})$. Note that in the diagram factorizable $\vert 0\rangle$ states are left out for convenience.}
    \label{fig:gexample}
\end{figure}

Two simple genomes representing a 3- and a 4-spin network are given in Figure \ref{fig:gexample}, (a) and (b) respectively. In Figure \ref{fig:gexample}a), an initial excitation is injected at site A, $\vert\Psi(0)\rangle=\ket{100}_{ABC}$, and then, at each time within a chosen set, the fidelity is evaluated against the target state $\vert\Psi_{\text{target}}\rangle=\ket{001}_{ABC}$. This genome specifies two coupling energies with relative values of 500 (between A and B) and 500 (between B and C), thus representing a uniform chain.

In Figure \ref{fig:gexample}b), a more complicated example is presented. This genome represents a small spin network aiming to generate an entangled state between sites C and D, $\vert\Psi_{\text{target}}\rangle=\ket{00}_{AB}\otimes\frac{1}{\sqrt{2}}(\ket{10}_{CD}+\ket{01}_{CD})$, when a single excitation is injected at site A, $\vert\Psi(0)\rangle=\ket{1000}_{ABCD}$.

There are also some more niche genome features. A target time can be specified anywhere in the genome with syntax ``@12.40'' to force the algorithm to optimize the dynamics of the system to reach the target state at, for instance, $t_f=12.40/J_\text{max}$. This substantially increases the speed of the algorithm since only a single-point calculation corresponding to that specific time point is required, rather than evaluating the full dynamics (and searching) over a time window. Although this allows for faster optimisation, it removes any flexibility in the time and thus should only be used if the transfer time is known, or is to be specified rather than allowed to vary. As such, it is often useful to perform a short optimisation without this feature first, to see the time that the system naturally evolves towards, and then perform a subsequent optimisation specific to that time.

Non-uniform on-site energies can also be specified, achieved by including a repeated-letter coupling (such as $AA650$). Negative couplings are also allowed, which could be of interest when evaluating systems of different magnetic order, such as anti-ferromagnetic lattices. \cite{struck2011quantum} These are achieved through specifying a coupling with the letters in reverse-alphabetical order, such that $BA500$ represents $-500$. Such notation is used to keep genome length constant and concise. There also exists optional notation related to the visualisation of the genome, detailed in the Supporting Information (included here as Section \ref{supmat}).

\subsection{Fitness}

For the evaluation of the genomes one needs to define first a fitness function which takes a genome string as an input and returns its fitness (normalized to be a number between $0$ and $100$), indicating how well such a genome satisfies the protocol or device requirements. This function combines various factors such as the maximum fidelity between the evolving state and the target state, along with the time at which that state is reached, all scaled by customisable parameters.  

We chose the function to be exponential in order to give any genome which is mutated positively a more significant boost in fitness, such that a small increase in maximum fidelity results in a much greater fitness score, and long times taken to reach the maximum fidelity are penalised. The overall equation for this fitness function is given as

\begin{equation}
    f(F_\text{max}, t_f) = 100~ \exp{\left( a (F_\text{max}-1)\right )} ~\exp{\left (b t_f J_\text{max} \right )},
    \label{eqn:fitness}
\end{equation}

where $F_\text{max}$ is the maximum fidelity, $t_f$ is the time, in units of $1/J_{\text{max}}$, to reach $F_\text{max}$, and $a$ and $b$ are suitably chosen scaling factors. For the fidelity, the fitness function clearly grows exponentially with $F_\text{max}$, but once $1 - F_\text{max}$ decreases below $1/a$ this strong $F_\text{max}$-dependence flattens off. We have therefore found it effective to use $a=10$, since for many systems a fidelity greater than 90\% is then reached quickly, with the remainder of the optimisation spent on fine-tuning the system $F_\text{max}$ towards unity, or 100\%. For the time $t_f$, with $b<0$ the second exponential term in the fitness function clearly encourages vanishingly small values of $t_f$, but doesn't significantly penalise the time until $t_f J_{\text{max}} \sim 1/|b|$. We have therefore found it effective to use $b=-0.001$ for the calculations presented here. Increasing $|b|$, for example to $b=-0.01$, would place more emphasis on short $t_f$ in the fitness function, compared to the fidelity behaviour.

In any given run, the maximum fidelity is found by searching the dynamics of the system over a search window, by default between $0$ and $20/J_\text{max}$ divided into $100$ increments. The time-window width, $20/J_\text{max}$, was chosen to allow the dynamics to reach peak fidelity in spin networks of the size here considered. This time-window is in fact about one order of magnitude larger than $1/J_\text{max}$, the typical time an excitation would take to tunnel through a link with  maximum coupling.

Time window and number of increments were chosen to provide a balance between search resolution and performance, and can be easily changed if larger or more precise search regions are needed, perhaps for transfer over very long chains or for rapidly fluctuating fidelities, respectively. For all of the examples in this paper, however, this range is able to capture the maximum fidelity without requiring too many increments.

A fitness score of $100$ for a spin-chain designed as a state-transfer device would thus mean that the transfer has been unrealistically achieved with zero waiting time ($t_{f}=0$) and perfect fidelity ($F(t_f)=1$), whilst a value approaching $0$ would suggest either no information transfer at all or that it takes so long in time that it is not an effective solution. This fitness is then used to determine the likelihood that the features contained within a certain genome will continue through the generations.

\subsection{Crossover and Mutation}

In order to create the next generation, genomes from the previous generation are selected with a probability proportional to their fitness score. When two genomes are selected, they are combined in a process known as crossover. In this particular implementation, crossover involves iterating over the number of characters in the parent genomes and for each genome position randomly choosing (with equal probability) one of the parents from which to take the character, as shown visually in the crossover panel of Figure \ref{fig:flow}. Note that since both genomes share the same letter order, it is only the coupling values which change.

This new genome, also referred to as the child, is then ``mutated'' by increasing or decreasing one of its couplings by a random integer less than or equal to the maximum mutation size, $\mu$, generated uniformly. The new coupling is capped between $0$ and the highest possible coupling for that genome, unless negative couplings are explicitly allowed. This $\mu$ begins at some initial value, $\mu_i$, and is linearly decreased to some specified final value, $\mu_f$, as the generations continue to allow the algorithm to make more specific changes after initially covering a very wide search space. $\mu_i$ is by default 20\% of the maximum possible coupling, such that for a 3 digit genome each mutation could initially change by up to $20\%$ of $999$: meaning $\mu_i=200$. However, this should be changed to be higher or lower if a system requires more or less extreme changes, respectively. $\mu_f$ is set to unity by default, but should also be increased if $\mu$ should remain higher throughout the optimisation. New genomes are generated in this manner until an entirely new generation is created to replace the old one.

The overall process then repeats for a large number of generations, with each iteration resulting in an increased average fitness score until either a target fitness is reached or the program reaches some maximum elapsed iteration. Unless otherwise stated, optimisations were run for 200 generations, each containing 1024 genomes. These values were chosen as they allow sufficient time/diversity for most systems of these sizes. Note that, for optimum parallel performance, the number of genomes should be chosen as a multiple of the number of CPU cores used to perform the optimisation; thus powers of two work well.

\section{Applications}\label{sec:results}

We now provide examples of the application of the aforementioned algorithm for the design of different quantum devices. We will consider the on-site energies $\epsilon_i$ to be uniform and scaled to zero for the examples presented here.

\subsection{Quantum State Distribution}

As with classical computers and conventional data, a quantum computer processor requires quantum networks to be able to transmit quantum data between registers. Clearly the use of photonics for such short range communication presents some drawbacks: quantum computer hardware is generally built out of static matter qubits (e.g. superconductors, ion traps or quantum dots) and the use of photons would imply the conversion of the matter qubit state into states of light and vice-versa, a costly process for such short distances. Instead, the idea of using linear spin chains as quantum data buses has attracted significant interest, \cite{bose2003,bose2007,christandl2004,christandl2005,kay2010} motivated by the possibility of building a ``wire" with the same type of solid-state qubit as the rest of the hardware, to avoid conversion between different forms of qubits. One of the most well-known methods for doing this uses the natural dynamics of the system by engineering the spin-spin interactions of a one-dimensional chain. \cite{christandl2004, christandl2005} Using only the natural dynamics implies that once the interactions are set, no further external control is required. We have used such existing results to both verify the accuracy of the algorithm (refer to the Supporting Information, included here as Section \ref{supmat}) and to test the algorithm's scaling (discussed in section \ref{sec:scaling}).

\begin{figure}
    \centering
    \includegraphics[width=0.9\linewidth]{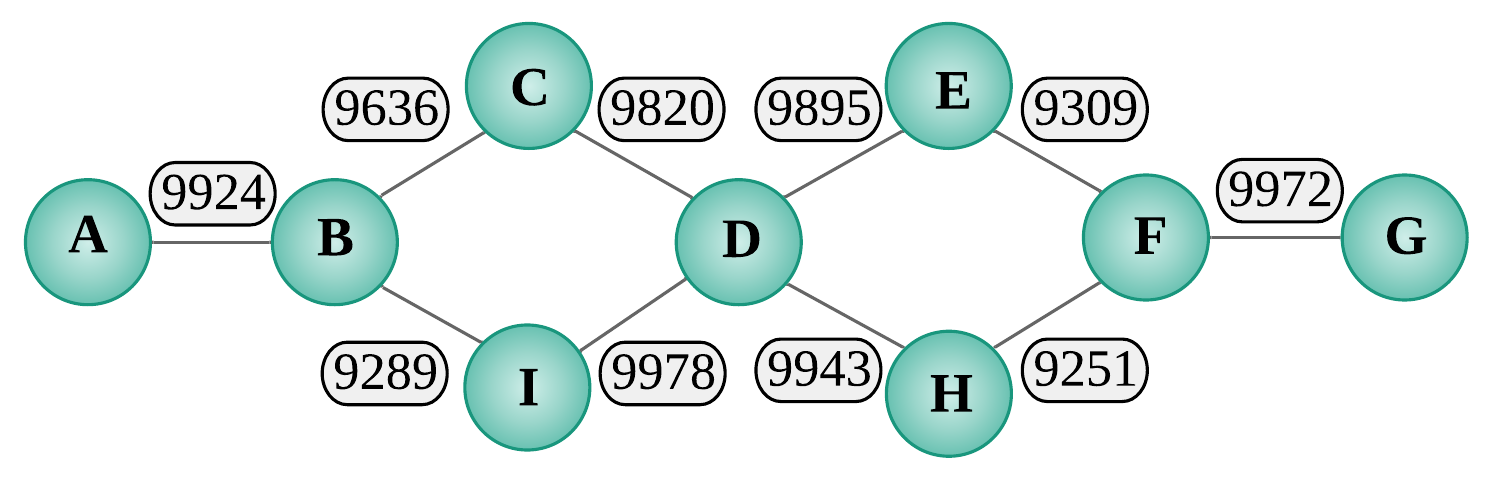}
    \caption{A network of spins arranged in a shoelace pattern optimized for speed and quantum state transfer. This maps $\ket{1}_A$ to $\ket{1}_G$ at time $t_f=3.8/J_\text{max}$ with $99.7\%$ fidelity.}
    \label{fig:shoelace}
\end{figure}

In this section, we focus on optimisation of more complex, non-trivial networks for quantum state transfer and entanglement generation, for which our algorithm becomes more interesting. Here a simple ``shoelace'' network was chosen as an example topology, initialized uniformly. When optimized for state transfer between sites $A$ and $G$, it resulted in a transfer time of $t_f=3.8/J_\text{max}$ with $99.70\%$ fidelity, with the resultant structure shown in Figure \ref{fig:shoelace}. This is $32\%$ faster than a 7-site linear chain engineered with previously-known perfect state transfer capabilities, \cite{christandl2005} and is attained by adding two extra nodes that modify the topology away from a linear chain. The identification of faster structures highlights the potential for this method to create improved spin-channels between quantum computing components.

Changing the parameters of the fitness function would result in a different result being converged upon, allowing flexibility depending on the physical implementation, e.g. taking into account factors like the decoherence time. For instance, in the example above, an emphasis on shorter $t_f$ was requested by changing the value of $b$ (from Equation \ref{eqn:fitness}) to $-1000$. Running with the standard value of $b=-0.001$ places more emphasis on the system's fidelity, resulting in a transfer time of $4.8/J_\text{max}$, at a fidelity of $99.8\%$.

\setlength{\tabcolsep}{1.0em}
\renewcommand{\arraystretch}{1.7}
\begin{table*}
\centering
\begin{tabular}{ c  c  c  c  c  }
\hline
Task (Genome Notation) & Initial & Target & Fidelity & Time $\cdot J_\text{max}$  \\
\hline
$ \braket{A|G}@5.0 $ & $ \ket{1}_{A} $  & $ \ket{1}_{G} $  & $ 99.7\% $ & $ 5.0 $ \\
$ \braket{A|G}@8.0 $ & $ \ket{1}_{A} $  & $ \ket{1}_{G} $  & $ 99.8\% $ & $ 8.0 $ \\
$ \braket{A|E} $ & $ \ket{1}_{A} $  & $ \ket{1}_{E} $  & $ 99.9\% $ & $ 3.8 $ \\
$ \braket{C|A+G} $ & $ \ket{1}_{C} $  & $ \frac{1}{\sqrt{2}}(\ket{10}_{AG}+\ket{01}_{AG}) $  & $ 99.8\% $ & $ 3.0 $ \\
$ \braket{C|A-G} $ & $ \ket{1}_{C} $  & $ \frac{1}{\sqrt{2}}(\ket{10}_{AG}-\ket{01}_{AG}) $   & $ 99.9\% $ & $ 3.4 $ \\
$ \braket{C+H|I+E} $ & $ \frac{1}{\sqrt{2}}(\ket{10}_{CH}+\ket{01}_{CH}) $  & $ \frac{1}{\sqrt{2}}(\ket{10}_{IE}+\ket{01}_{IE}) $  & $ 99.9\% $ & $ 4.8 $ \\
$ \braket{AB|FG} $ & $ \ket{11}_{AB} $  & $ \ket{11}_{FG} $  & $ 90.0\% $ & $ 4.4 $ \\
\hline
\end{tabular}
\caption{The various test optimisations performed on the shoelace topology (see Figure \ref{fig:shoelace}). This shows how our method is capable of finding solutions to many different tasks. Note each of these represents a different coupling scheme optimized for that purpose, not a single coupling scheme achieving all tasks. Each of these optimisations used standard parameters, with runtimes taking between 5 and 30 seconds using 8 cores. \cite{computer}}
\label{tbl:tests}
\end{table*}

To further show the adaptability of this method we perform optimisations on this same topology for various quantum information tasks, including examples of quantum state transfer tailored to occur at some chosen time, entanglement generation between arbitrary sites and multi-excitation transfer. A summary of such tests is given in Table \ref{tbl:tests}, all showing high fidelities. We note that, for tests in which specific times have not been specified, shorter transfer times may be requested at the cost of fidelity.

\begin{figure}
    \centering
    \includegraphics[width=1.0\linewidth]{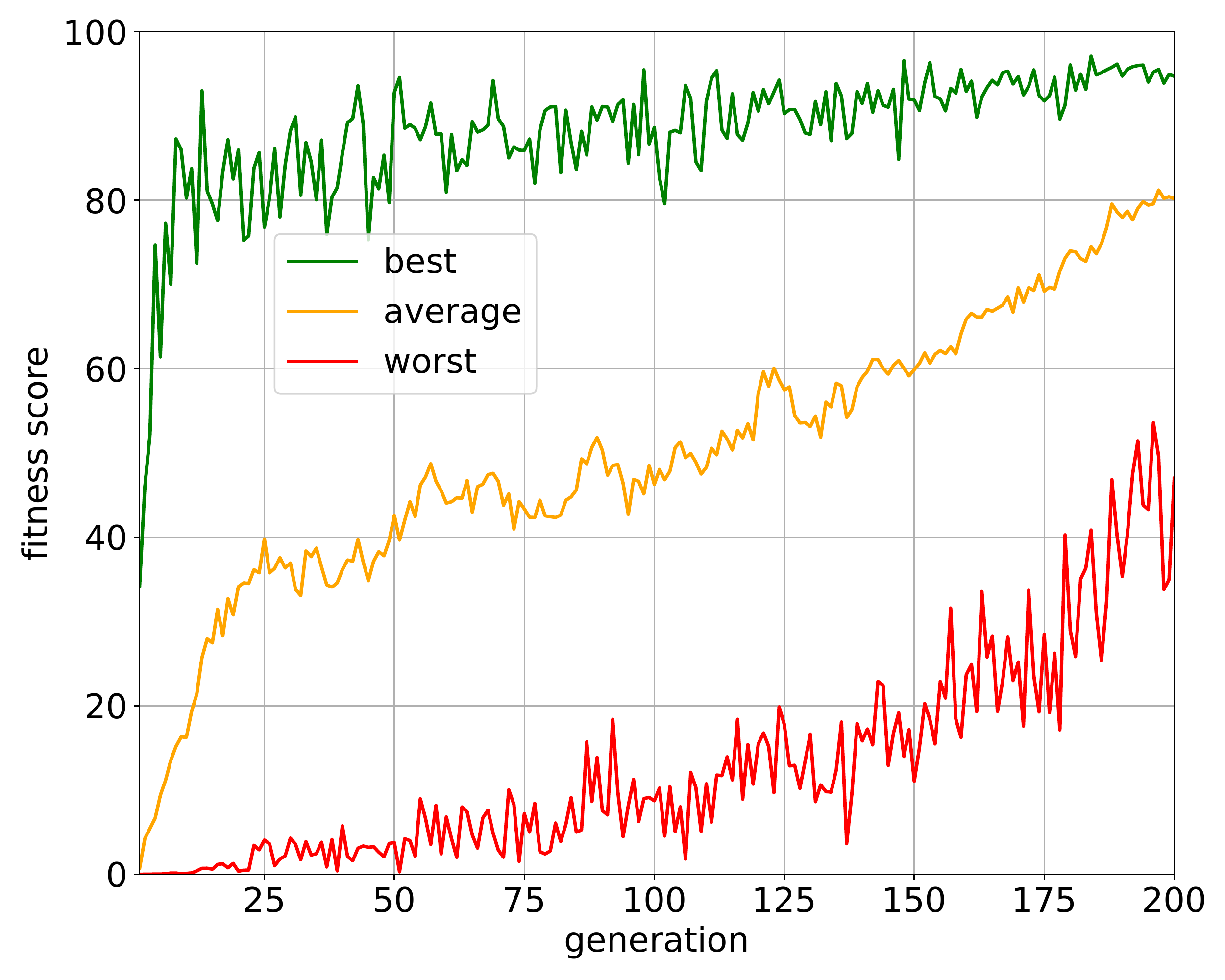}
    \caption{Worst, average and best fitness scores $f(F_\text{max},t_f)$ for each generation when optimising a shoelace network for quantum state transfer, generating the result shown in Figure \ref{fig:shoelace}. The size of mutations is reduced each generation, resulting in smaller, more precise, changes in fitness, except for the worst fitness scores, which are the product of more diverse genomes. Running all 200 generations has a runtime of around 5 seconds using 8 cores. \cite{computer}}
    \label{fig:geneticShoelace}
\end{figure}

The best, average and worst fitness scores for each generation during the optimisation process leading to the values in Figure \ref{fig:shoelace} are displayed in Fig~\ref{fig:geneticShoelace}, which shows how few generations are required for this method to reach a high fitness score. Note that here the optimisation was stopped after the default maximum number of generations (200) as a demonstration, but could have been stopped much sooner and still retained fast and high fidelity state transfer. Corresponding graphs for the optimisation of other systems are given in the Supporting Information (included here as Section \ref{supmat}). Importantly, although a simple method could be to take only the best genomes for each generation, the best fitness may then become trapped at a local maximum, whilst worse solutions may eventually reach an overall higher fitness if allowed to evolve down their path.

\subsection{Design of a Quantum Gate}

\setlength{\tabcolsep}{1.0em}
\renewcommand{\arraystretch}{1.7}
\begin{table*}
\centering
\begin{tabular}{ c  c  c  c  c  c }
\hline
Initial &  Final & 4 s.f. & 3 s.f. & 2 s.f. & 1 s.f. \\
\hline
$ \ket{01}_{RA} $ & $ \ket{01}_{SF} $ & 99.8\% & 99.9\% & 99.8\% & 89.0\% \\
$ \ket{10}_{RA} $ & $\ket{10}_{SF}$ & 99.9\% & 99.9\% & 99.9\% & 91.9\% \\
$ \ket{11}_{RA} $ & $-\ket{11}_{SF}$ & 99.8\% & 99.8\% & 99.6\% & 83.8\% \\
$ \frac{1}{2} (\ket{00}_{RA}+\ket{01}_{RA} $ & $ \frac{1}{2} (\ket{00}_{SF}+\ket{01}_{SF} $ &\multirow{2}{*}{99.8\%} & \multirow{2}{*}{99.8\%} & \multirow{2}{*}{99.8\%} & \multirow{2}{*}{89.0\%}\\
 $ ~~~+\ket{10}_{RA}+\ket{11}_{RA}) $ & $ ~~~+\ket{10}_{SF}-\ket{11}_{SF}) $  \\
\hline
\end{tabular}
\caption{Truth table showing the fidelities when the controlled-phase-gate genome is evaluated with different input injections and with various levels of genome precision, given as the number of significant figures (s.f.) used per coupling. Here by ``$n$ significant figures'' we mean rounding each coupling to the nearest $10^{4-n}$, e.g. $1432\to 1000$ for $n=1$. Note that all sites are assumed to have no excitation unless otherwise specified (such that $\ket{0}$ states of non-relevant sites are omitted for clarity). The system is tailored so that each of these outputs is achieved at the same time of $12.4/J_\text{max}$.}
\label{tbl:phase}
\end{table*}

Whilst transferring quantum information quickly and reliably is one of the most popular uses of spin networks, when it comes to designing quantum hardware there may be situations where it would significantly aid computation if a quantum gate could be applied to the information as it is being transferred. To do this with large spin networks one would need to find suitable tuning of the numerous parameters, something which would be difficult to achieve analytically. We thus identify such an example as one of the most appropriate use cases for our proposed method.

\begin{figure}
    \centering
    \includegraphics[width=0.9\linewidth]{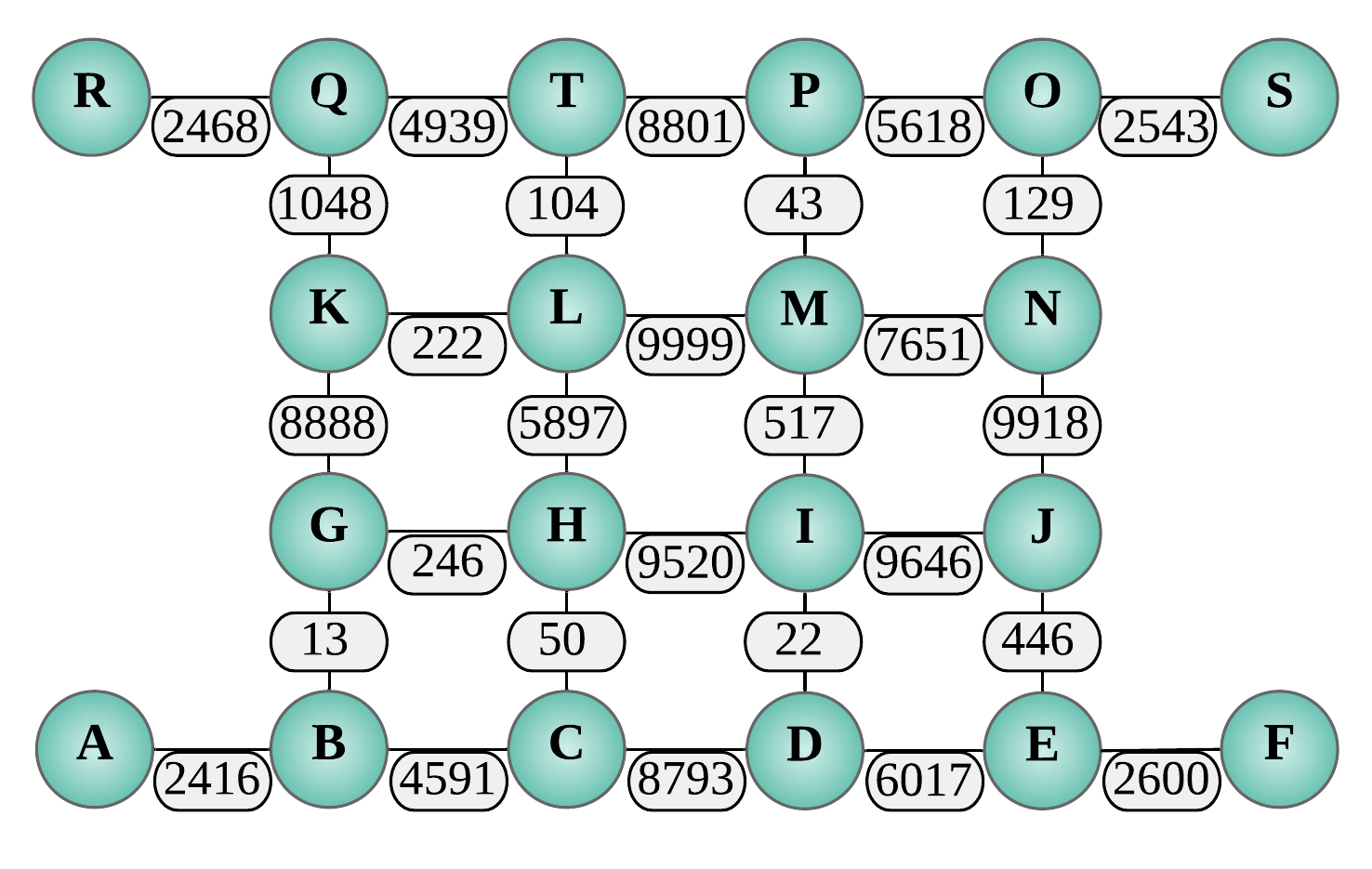}
    \caption{A network optimized to perform a controlled phase gate on the two qubits, and mapping the state $ \frac{1}{2} (\ket{00}_{RA}+\ket{10}_{RA}+\ket{01}_{RA}+\ket{11}_{RA}) $ to $ \frac{1}{2} (\ket{00}_{SF}+\ket{10}_{SF}+\ket{01}_{SF}-\ket{11}_{SF}) $ with 99.8\% fidelity at time $t_f=12.40/J_\text{max}$.}
    \label{fig:phase}
\end{figure}

In the following test we consider a $4\times 4$ grid topology as a ``blank canvas" for larger systems, plus two input and two output spins, with the aim of engineering the system to perform a controlled-Z gate on two qubits (see Table \ref{tbl:phase}). In Figure \ref{fig:phase} we draw the topology along with the optimized couplings: the gate is to be applied between qubits R and A, with the result being output at sites S and F. A key challenge in this application is the constraint that all input/output operations in the corresponding truth table (first three lines of Table \ref{tbl:phase}) must be achieved in the same output time, and, of course, all with high fidelity.

The genetic algorithm was capable of identifying the tuning outlined in Figure \ref{fig:phase} with the device able to perform a controlled-Z gate, allowing for the initial product state

\begin{equation*}
\vert \Psi (0)\rangle = \frac{1}{2}(\vert 00\rangle_{RA}+\vert 01\rangle_{RA}+\vert 10\rangle_{RA}+\vert 11\rangle_{RA}),
\end{equation*}

to be mapped to the approximate final state of

\begin{equation*}
\vert \Psi (t_f)\rangle \approx \frac{1}{2}(\vert 00\rangle_{SF}+\vert 01\rangle_{SF}+\vert 10\rangle_{SF}-\vert 11\rangle_{SF})
\end{equation*}

with $99.8\% $ fidelity and a transfer time of $t_f=12.40/J_\text{max}$.

Importantly, this network is shown to retain the high $\approx 99.8\%$ fidelity even as the number of digits used in the genome is approximated from 4 to 3 and even 2 significant figures (s.f.), which would allow tolerance when implementing such a network in the lab, as shown in Table \ref{tbl:phase}. 
The approximation is done such that, for example, the 2 s.f. couplings are the 4 s.f. couplings, but rounded to the nearest $100$ and then divided by $100$ (e.g. $23\to 0$, $2524\to 25$ etc.). The fidelity for the genome with these 2 s.f. couplings is then evaluated and reported in Table \ref{tbl:phase}. 
Even when the couplings are rounded up just to the nearest 1000 (right-end column in Table 2), the resulting approximated solution still retains a very high fidelity for the requested task. This implies that the minimum requirement for experiments is quite modest, that is to be able to vary coupling energies between a set reference value ($J_{max}$) and a tenth of it, a modest requirement. While we cannot claim it to be a general result, we found that this robustness is shared by various other examples. This high tolerance would also suggest that to improve the performance of this particular network a change in the topology is needed, rather than simply increasing precision of the genome.

In this example, unlike the others, the two-excitation coupling term in Equation \ref{eqn:hamil} affects the results when $\alpha\neq 0$. This term helps to build a phase specific to subspaces containing at least two excitations, allowing the network for the controlled phase gate to be optimized to reach higher fidelities. Without such a term, this topology was able to reach at most $77\%$ fidelity, whilst optimising with $\alpha=0.141$ allowed for the $99.8\%$ fidelity result. In some physical implementations, this coupling term would correspond to a dipole-dipole interaction. A scaling factor of $\alpha=0.141$ with respect to each coupling $J_{ij}$ is then consistent with this second-order type of interaction.

\begin{figure}
    \centering
    \includegraphics[width=0.90\linewidth]{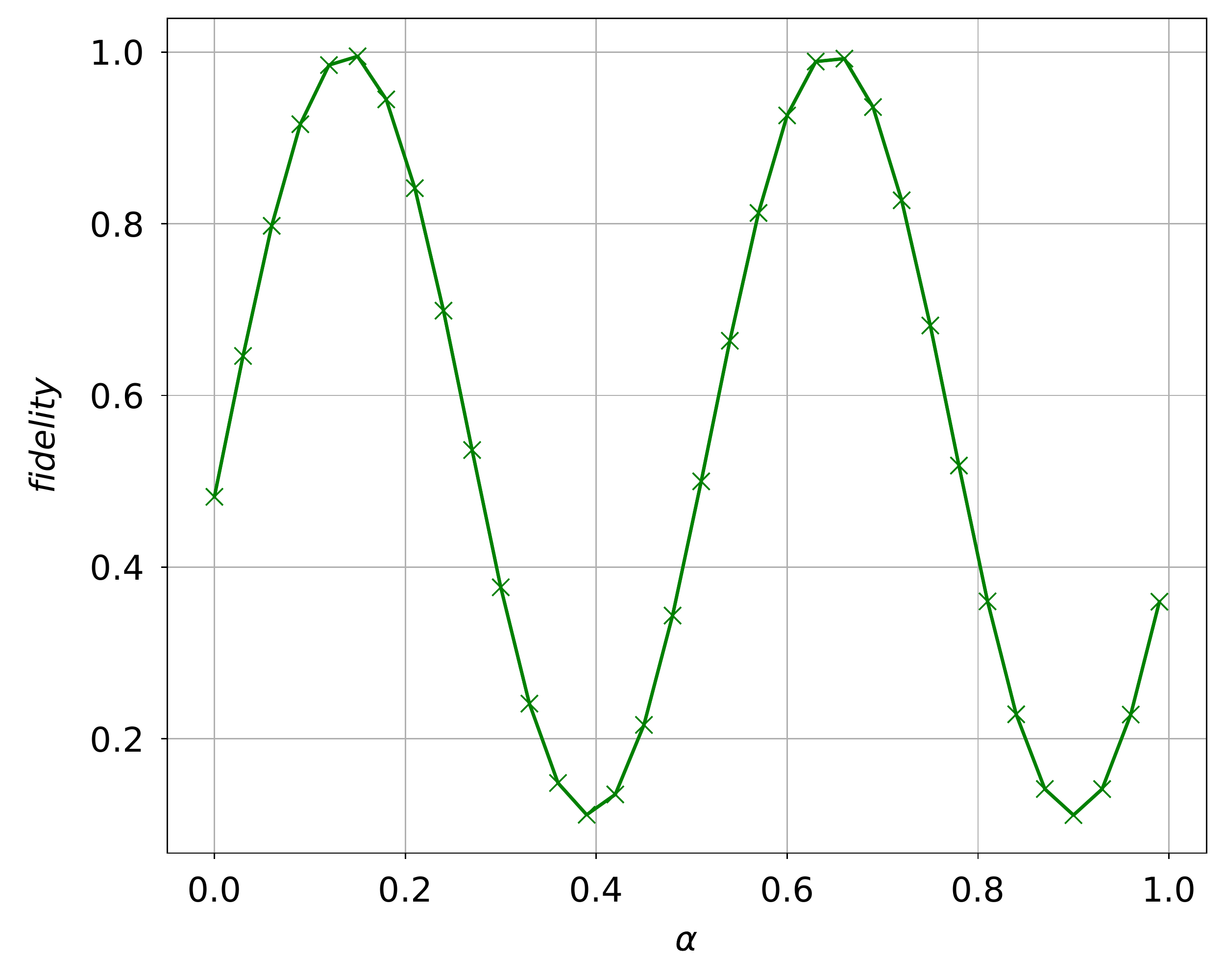}
    \caption{Fidelity $F(t_f)$ of the phase-gate vs $\alpha$. The fidelity shows sinusoidal behaviour with respect to $\alpha$. The first peak is reached at $\alpha=0.141$ with a fidelity of $99.8\%$.}
    \label{fig:alphaScale}
\end{figure}

Figure \ref{fig:alphaScale} shows how the fidelity is affected when the phase-gate coupling scheme is evaluated using different values of $\alpha$. It shows that the gate design is robust against small variation of $\alpha$ about its best value. The fidelity describes a sinusoidal variation with respect to $\alpha$, confirming that this term is responsible for the creation of a phase, and hence offering multiple choices of $\alpha$ for achieving best fidelity.

\section{Parallelisation and scaling}\label{sec:scaling}

An interesting property of genetic algorithms is that they can be highly parallelized. In our case, we have been able to efficiently parallelize our algorithm using a standard implementation of the Message Passing Interface (MPI). This allowed large networks with multiple excitations to be fully optimized within an hour, which otherwise would have taken a day. This is all done by distributing the evaluation of each generation between the CPU cores, providing embarrassingly-parallel speedup.

\begin{figure}[t]
    \centering
    \includegraphics[width=1.0\linewidth]{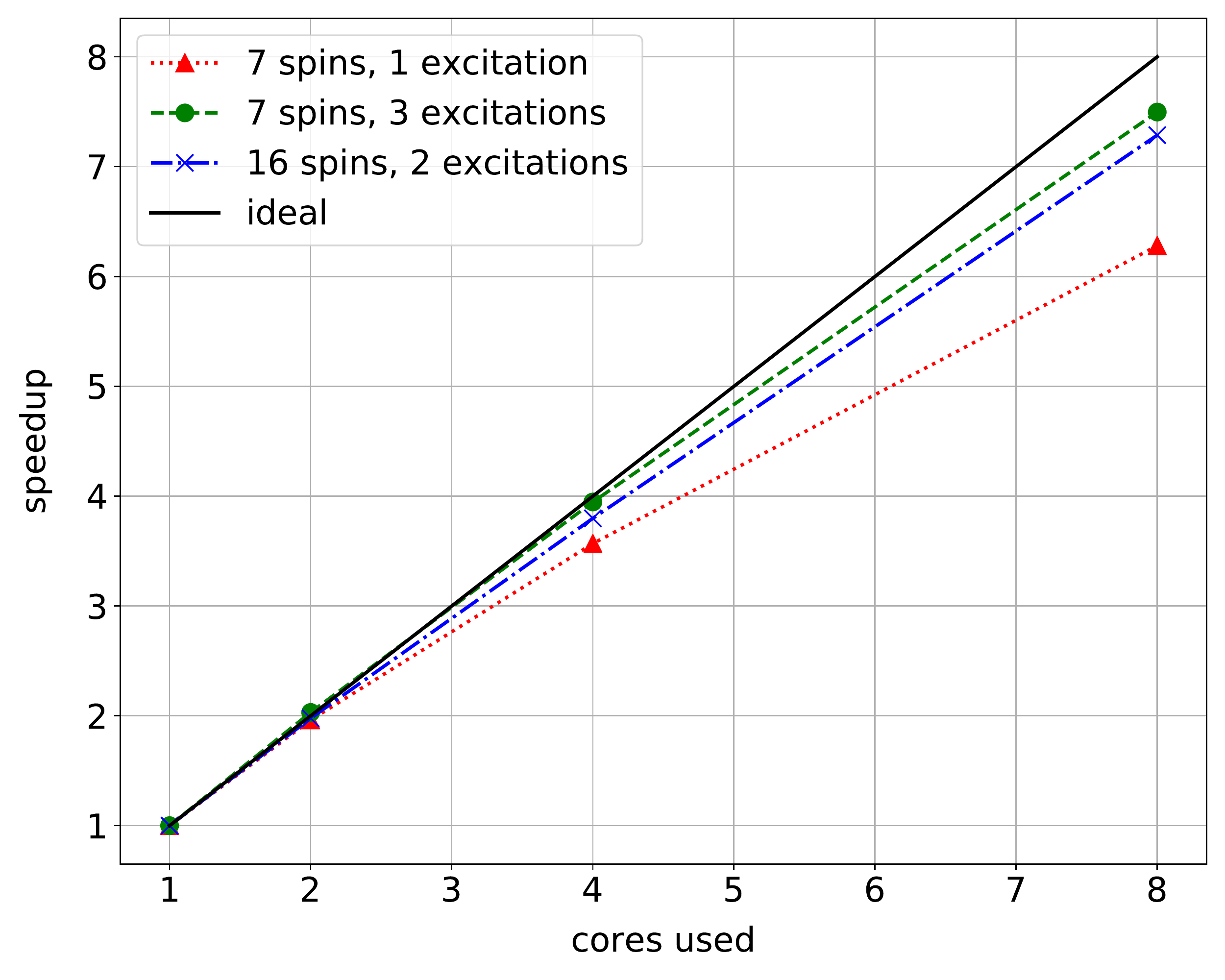}
    \caption{Scaling of the algorithm for a given system as the number of CPU cores used in parallel is increased. Here $\text{speedup}=t_1/t_n$, with $t_n$ the time for a certain optimization using $n$ cores. Three systems are used for demonstration: a linear chain of 7 sites using a single excitation, the same chain but with three excitations, and a 16 spin grid with two excitations. Each is optimized using 1024 genomes for 200 generations. Each point was averaged over 5 realisations \cite{computer}. Note how scaling becomes closer to the ideal case as the subspaces become larger, allowing more efficient CPU usage.}
    \label{fig:scalingParallel}
\end{figure}

The parallel performance of the code is shown in Figure \ref{fig:scalingParallel}, which shows how the time taken to optimize a system is reduced by a factor of approximately two using two cores, four with four cores etc., a concept known as parallel speedup, ideally an identity function. The optimisations used for this test were run for a fixed number of generations (here 200) to focus more on the efficiency of the algorithm than on the ease of optimising each given system. A positive feature is that the scaling is better for larger systems, since more time is spent evaluating each genome, a task done entirely in parallel, compared to smaller systems in which most of the time is spent on the more trivial serial operations, such as distributing/collecting the genomes between cores.

\begin{figure}
    \centering
    \includegraphics[width=1.0\linewidth]{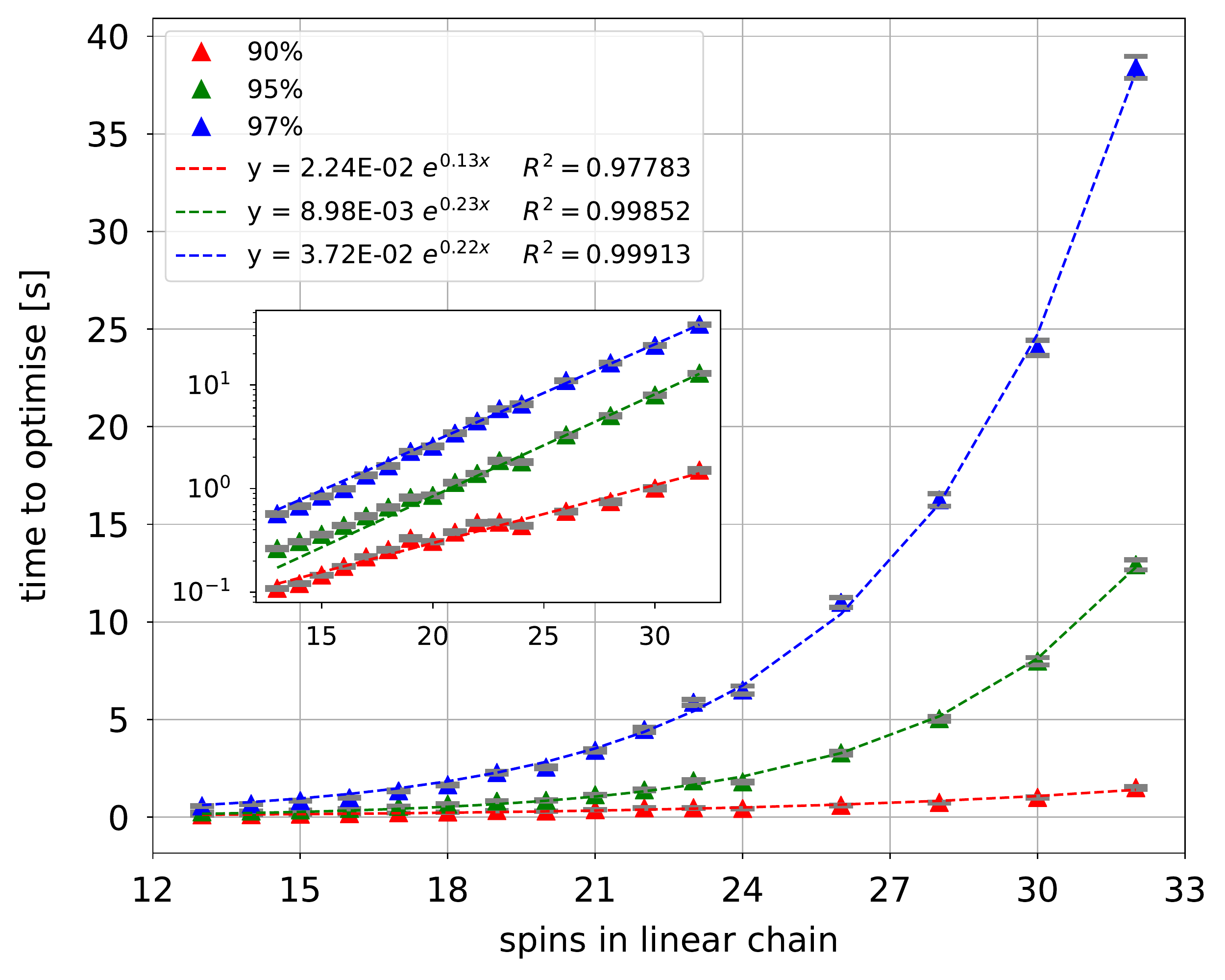}
    \caption{Scaling of the algorithm as the system size increases, specifically for a linear chain optimized to transfer fidelities of 90\%, 95\% and 97\%. Each point was averaged over 400 realisations, each done in parallel using 4 cores \cite{computer2}. These optimizations all use the same parameters for a fair comparison, however, performance can be improved by tweaking parameters on a case-by-case basis.}
    \label{fig:scalingSites}
\end{figure}

We also consider how the algorithm scales as the systems become larger. This was done by extending a linear chain and timing how long was required to reach 90\%, 95\% and 97\% fidelity for an end-to-end state transfer. The results are given in Figure \ref{fig:scalingSites} and show that even for large systems of 32 qubits, and using just four cores, the optimisations are still performed in under a minute to a very high fidelity and in only a few seconds to good fidelity \cite{computer2}. All optimisations here use the same algorithm parameters (such as the number of genomes or the maximum mutation size) to allow for a fairer comparison. If improved performance is desired, these parameters should be manually tailored for each system.

For the system shown in Figure \ref{fig:scalingSites}, the time $t_{opt}$ required to optimise such a system is shown to scale exponentially with the system size, albeit with very small coefficients, e.g. $t_{opt}=4\times 10^{-2}\exp(0.22N)$ for $97\%$ fidelity and $N$ spins. This is as expected, since the search space grows exponentially with every added coupling (one coupling has 9999 possibilities, two have a total of $9999^2$, and so on). Further analysis on the algorithm's scaling is given in the Supporting Information (included here as Section \ref{supmat}), where the algorithm is run for a fixed number of generations (200), but for different numbers of excitations, to more directly show the effect of increasing the size of the Hilbert space.

Whilst we cannot compare the efficiency of our method with previous approaches due to a lack of benchmarks, we can study how it compares to a randomized search of the parameter space. Our results show that even for a modest 10 qubit linear chain (9 parameters to be optimized), the randomized approach reached $99\%$ fidelity only after 48 minutes of parallelized searching, while our genetic algorithm converges 5000 times faster for that same example. If we now move to a larger system, such as our proposed phase gate (28 parameters to be optimized), it would be completely infeasible to find appropriate solutions either randomly or analytically due to the sheer size of the parameter space. Also, our 5000 times speedup against the randomized approach is in contrast with the one obtained by another piece of research, \cite{moon2020} where their machine learning algorithm yields optimization times 180 times faster than the automated random search of the parameter space, although a direct comparison is difficult due to the different nature of the two problems.

\section{Conclusions}\label{sec:conclusion}

As quantum devices improve in terms of number of qubits and connectivity, the power of quantum computation gets an exponential boost. However, this comes with the challenge that an increased number of degrees of freedom brings to the tuning of parameters when engineering quantum devices. To overcome this, we propose a novel method based on evolutionary computation that is capable of efficiently finding appropriate solutions from within the large search space of possible parametrizations. This method exploits a genetic algorithm that we have specifically designed to optimize the different degrees of freedom of a spin network to perform any given quantum information task. We provide examples showing new network designs discovered by the algorithm, acting as a controlled phase gate on two qubits, an entangler, or allowing for fast and high fidelity quantum information transfer between arbitrary sites of a network. Such networks are shown to retain high fidelities even when the precision on the coupling energies is reduced, which would allow a margin for fabrication and control errors when implementing such devices experimentally. To demonstrate flexibility, we have used some example networks with non-trivial topologies, where multiple paths exist between points of interest in the network.


It is important to note that our method allows optimisation with a tailored fitness function as well as the inclusion of a set of optional and flexible parameters. This versatility provides the possibility of a variety of use cases depending on the specific experimental constraints. For example, when designing a quantum state transfer device it may be preferable to transfer information faster at the expense of a lower fidelity in cases where decoherence times are relatively short. Our algorithm can be programmed with such constraints, making our method a promising candidate to assist in the design and calibration of real NISQ devices. Due to the highly parallelisable nature and efficiency of genetic methods, the algorithm converges rapidly, often within only seconds, with convergence occurring approximately 5000 times faster than the equivalent randomized search for a 10-parameter network. It also scales well with an increasing number of spins and excitation subspaces, with trends that, even though exponential, present small coefficients: very good solutions are found after just a small number of generations and thus in a relatively short time, for example less than 40$s$ are needed for optimizing a 32-parameters spin chain to $97\%$ fidelity.

The method we propose can be easily extended to include additional terms in the Hamiltonian, or to different model Hamiltonians. Future research will involve testing the method against a larger set of initial topologies and quantum information tasks, such as optimising for different quantum gates. Although our code allows for the optimisation of the on-site energies, in this work it was decided to focus on optimising only the coupling values, to allow us to compare with more well known results. However, future investigations that also include on-site energies are clearly of interest, not only both from a theory and modelling perspective, but also with respect to physical implementations. Further investigations will also involve extending our methods to provide results bound by the experimental constraints, for example given by a specific quantum chip implementation. 

\medskip
\textbf{Acknowledgements} \par 

M.P.E would like to acknowledge support from the Japanese MEXT Quantum Leap Flagship Program (MEXT Q-LEAP) Grant Number JPMXS0118069605. This project was in part undertaken on the Viking Cluster, a high performance computing facility provided by the University of York. We are grateful for computational support from the University of York High Performance Computing service, Viking and the Research Computing team.

\medskip


\pagebreak

\bibliographystyle{unsrt}
\bibliography{refs}

\begin{thebibliography}{10}

\bibitem{preskill2012quantum}
John Preskill.
\newblock Quantum computing and the entanglement frontier, 2012.

\bibitem{Arute2019}
Frank Arute, Kunal Arya, Ryan Babbush, Dave Bacon, Joseph~C Bardin, Rami
  Barends, Rupak Biswas, Sergio Boixo, Fernando~GSL Brandao, David~A Buell,
  et~al.
\newblock Quantum supremacy using a programmable superconducting processor.
\newblock {\em Nature}, 574(7779):505--510, 2019.

\bibitem{Liao2017}
Sheng-Kai Liao, Wen-Qi Cai, Wei-Yue Liu, Liang Zhang, Yang Li, Ji-Gang Ren,
  Juan Yin, Qi~Shen, Yuan Cao, Zheng-Ping Li, Feng-Zhi Li, Xia-Wei Chen, Li-Hua
  Sun, Jian-Jun Jia, Jin-Cai Wu, Xiao-Jun Jiang, Jian-Feng Wang, Yong-Mei
  Huang, Qiang Wang, Yi-Lin Zhou, Lei Deng, Tao Xi, Lu~Ma, Tai Hu, Qiang Zhang,
  Yu-Ao Chen, Nai-Le Liu, Xiang-Bin Wang, Zhen-Cai Zhu, Chao-Yang Lu, Rong Shu,
  Cheng-Zhi Peng, Jian-Yu Wang, and Jian-Wei Pan.
\newblock {Satellite-to-ground quantum key distribution}.
\newblock {\em Nature}, 549(7670):43--47, 2017.

\bibitem{Wehnereaam9288}
Stephanie Wehner, David Elkouss, and Ronald Hanson.
\newblock Quantum internet: A vision for the road ahead.
\newblock {\em Science}, 362(6412), 2018.

\bibitem{Kurizki_2015}
Gershon Kurizki, Patrice Bertet, Yuimaru Kubo, Klaus Mølmer, David Petrosyan,
  Peter Rabl, and Jörg Schmiedmayer.
\newblock Quantum technologies with hybrid systems.
\newblock {\em Proceedings of the National Academy of Sciences},
  112(13):3866–3873, Mar 2015.

\bibitem{nikoBook}
Georgios~M Nikolopoulos, Igor Jex, et~al.
\newblock {\em Quantum State Transfer and Network Engineering}.
\newblock Springer, 2014.

\bibitem{christandl2004}
M.~Christandl, N.~Datta, A.~Ekert, and A.~J. Landahl.
\newblock Perfect state transfer in quantum spin networks.
\newblock {\em Phys. Rev. Lett.}, 92(18):187902, May 2004.

\bibitem{niko2004}
G.~M. Nikolopoulos, D.~Petrosyan, and P.~Lambropoulos.
\newblock Electron wavepacket propagation in a chain of coupled quantum dots.
\newblock {\em J. Phys. Condens. Matter}, 16(28):4991, Jul 2004.

\bibitem{wang2011all}
Yaoxiong Wang, Feng Shuang, and Herschel Rabitz.
\newblock All possible coupling schemes in xy spin chains for perfect state
  transfer.
\newblock {\em Physical Review A}, 84(1):012307, 2011.

\bibitem{karbach2005}
P.~Karbach and J.~Stolze.
\newblock Spin chains as perfect quantum state mirrors.
\newblock {\em Phys. Rev. A}, 72(3):030301, Sep 2005.

\bibitem{vinet2011}
L.~Vinet and A.~Zhedanov.
\newblock How to construct the spin chains with perfect state transfer.
\newblock {\em Phys. Rev. A}, 85(1):012323, Jan 2012.

\bibitem{estarellas2016}
Marta~P. Estarellas, Irene D'Amico, and Timothy~P. Spiller.
\newblock Topologically protected localised states in spin chains.
\newblock {\em Scientific Reports}, 7:42904, Feb 2017.

\bibitem{Andrea2016}
Andrea Blanco-Redondo, Imanol Andonegui, Matthew~J. Collins, Gal Harari, Yaakov
  Lumer, Mikael~C. Rechtsman, Benjamin~J. Eggleton, and Mordechai Segev.
\newblock Topological optical waveguiding in silicon and the transition between
  topological and trivial defect states.
\newblock {\em Phys. Rev. Lett.}, 116:163901, 2016.

\bibitem{Longhi2019}
Stefano Longhi.
\newblock Topological pumping of edge states via adiabatic passage.
\newblock {\em Phys. Rev. B}, 99:155150, Apr 2019.

\bibitem{damico2007_2}
I.~D'Amico, B.~W. Lovett, and T.~P. Spiller.
\newblock Freezing distributed entanglement in spin chains.
\newblock {\em Phys. Rev. A}, 76(3):030302, Sep 2007.

\bibitem{spiller2007}
T.~P. Spiller, I.~D'Amico, and B.~W. Lovett.
\newblock Entanglement distribution for a practical quantum-dot-based quantum
  processor architecture.
\newblock {\em New J. Phys.}, 9(1):20, Jan 2007.

\bibitem{estarellas2017robust}
Marta~P Estarellas, Irene D'Amico, and Timothy~P Spiller.
\newblock Robust quantum entanglement generation and generation-plus-storage
  protocols with spin chains.
\newblock {\em Phys. Rev. A}, 95(4):042335, 2017.

\bibitem{Apollaro2019}
Tony J.~G. Apollaro, Guilherme M.~A. Almeida, Salvatore Lorenzo, Alessandro
  Ferraro, and Simone Paganelli.
\newblock Spin chains for two-qubit teleportation.
\newblock {\em Phys. Rev. A}, 100:052308, Nov 2019.

\bibitem{ronke2011knitting}
R~Ronke, I~D’Amico, and TP~Spiller.
\newblock Knitting distributed cluster-state ladders with spin chains.
\newblock {\em Physical Review A}, 84(3):032308, 2011.

\bibitem{tserkovnyak2011universal}
Yaroslav Tserkovnyak and Daniel Loss.
\newblock Universal quantum computation with ordered spin-chain networks.
\newblock {\em Physical Review A}, 84(3):032333, 2011.

\bibitem{landahl2004information}
Andrew Landahl, Matthias Christandl, Nilanjana Datta, and Artur Ekert.
\newblock Information processing in quantum spin systems.
\newblock In {\em AIP Conference Proceedings}, volume 734, pages 215--218.
  American Institute of Physics, 2004.

\bibitem{Eiben2015}
Agoston~E Eiben and Jim Smith.
\newblock {From evolutionary computation to the evolution of things}.
\newblock {\em Nature}, 521(7553):476--482, 2015.

\bibitem{Eiben2003}
Agoston~E Eiben, James~E Smith, et~al.
\newblock {\em Introduction to evolutionary computing}.
\newblock Springer.

\bibitem{hornby2006automated}
Gregory Hornby, Al~Globus, Derek Linden, and Jason Lohn.
\newblock Automated antenna design with evolutionary algorithms.
\newblock In {\em Space 2006}, page 7242. 2006.

\bibitem{evans2017aerodynamic}
B~Evans and SP~Walton.
\newblock Aerodynamic optimisation of a hypersonic reentry vehicle based on
  solution of the boltzmann--bgk equation and evolutionary optimisation.
\newblock {\em Applied Mathematical Modelling}, 52:215--240, 2017.

\bibitem{Yao1999}
{Xin Yao}.
\newblock Evolving artificial neural networks.
\newblock {\em Proceedings of the IEEE}, 87(9):1423--1447, 1999.

\bibitem{Schmidt2009}
Michael Schmidt and Hod Lipson.
\newblock Distilling free-form natural laws from experimental data.
\newblock {\em Science}, 324(5923):81--85, 2009.

\bibitem{whitley1994genetic}
Darrell Whitley.
\newblock A genetic algorithm tutorial.
\newblock {\em Statistics and computing}, 4(2):65--85, 1994.

\bibitem{dominguez2015quantum}
Francisco Dom{\'\i}nguez-Serna and Fernando Rojas.
\newblock Quantum control using genetic algorithms in quantum communication:
  superdense coding.
\newblock In {\em Journal of Physics: Conference Series}, volume 624, page
  012009, 2015.

\bibitem{moon2020}
H.~Moon, D.~T. Lennon, J.~Kirkpatrick, N.~M. van Esbroeck, L.~C. Camenzind,
  Liuqi Yu, F.~Vigneau, D.~M. Zumb{\"u}hl, G.~A.~D. Briggs, M.~A. Osborne,
  D.~Sejdinovic, E.~A. Laird, and N.~Ares.
\newblock Machine learning enables completely automatic tuning of a quantum
  device faster than human experts.
\newblock {\em Nature Communications}, 11(1):4161, 2020.

\bibitem{esbroeck2020}
NM~van Esbroeck, DT~Lennon, H~Moon, V~Nguyen, F~Vigneau, LC~Camenzind, L~Yu,
  DM~Zumb{\"u}hl, GAD Briggs, Dino Sejdinovic, et~al.
\newblock Quantum device fine-tuning using unsupervised embedding learning.
\newblock {\em arXiv preprint arXiv:2001.04409}, 2020.

\bibitem{ronke2011effect}
R~Ronke, TP~Spiller, and I~D’Amico.
\newblock Effect of perturbations on information transfer in spin chains.
\newblock {\em Physical Review A}, 83(1):012325, 2011.

\bibitem{thede2004introduction}
Scott~M Thede.
\newblock An introduction to genetic algorithms.
\newblock {\em Journal of Computing Sciences in Colleges}, 20(1):115--123,
  2004.

\bibitem{struck2011quantum}
Julian Struck, Christoph {\"O}lschl{\"a}ger, R~Le~Targat, Parvis Soltan-Panahi,
  Andr{\'e} Eckardt, Maciej Lewenstein, Patrick Windpassinger, and Klaus
  Sengstock.
\newblock Quantum simulation of frustrated classical magnetism in triangular
  optical lattices.
\newblock {\em Science}, 333(6045):996--999, 2011.

\bibitem{bose2003}
S.~Bose.
\newblock Quantum communication through an unmodulated spin chain.
\newblock {\em Phys. Rev. Lett.}, 91(20):207901, Nov 2003.

\bibitem{bose2007}
S.~Bose.
\newblock Quantum communication through spin chain dynamics: an introductory
  overview.
\newblock {\em Contemp. Phys.}, 48(1):13, Jan 2007.

\bibitem{christandl2005}
M.~Christandl, N.~Datta, T.~C. Dorlas, A.~Ekert, A.~Kay, and A.~J. Landahl.
\newblock Perfect transfer of arbitrary states in quantum spin networks.
\newblock {\em Phys. Rev. A}, 71(3, Part A):032312, Mar 2005.

\bibitem{kay2010}
A.~Kay.
\newblock {\em Int. J. Quantum Inf.}, 8:641, 2010.

\bibitem{computer}
Simulated with a desktop computer with an AMD Ryzen 7 3700X.

\bibitem{computer2}
Simulated on a laptop with an Intel(R) Core(TM) i7-6700HQ CPU.

\bibitem{burgarth2005}
D.~Burgarth and S.~Bose.
\newblock Conclusive and arbitrarily perfect quantum-state transfer using
  parallel spin-chain channels.
\newblock {\em Phys. Rev. A}, 71(5):052315, May 2005.

\bibitem{osborne2004}
T.~J. Osborne and N.~Linden.
\newblock Propagation of quantum information through a spin system.
\newblock {\em Phys. Rev. A}, 69(5):052315, May 2004.

\bibitem{kostak2007}
V.~Kostak, G.~M. Nikolopoulos, and I.~Jex.
\newblock Perfect state transfer in networks of arbitrary topology and coupling
  configuration.
\newblock {\em Phys. Rev. A}, 75(4):042319, Apr 2007.

\end{thebibliography}

\pagebreak

\twocolumn[\section*{\centering Supporting Information \vspace{1.5em}}]
\label{supmat}

\subsection{Testing Against Known Results}

The use of chains of matter qubits as data buses requires a method to transport an arbitrary quantum state from one site to another.  Such quantum state transfer can be enabled with high fidelity through many strategies \cite{burgarth2005,osborne2004,kostak2007,karbach2005}. One of the most popular approaches is to utilise natural dynamics of the system by engineering the spin-spin interactions of a one-dimensional chain \cite{christandl2004,niko2004}.

\begin{figure}[h]
    \centering
    \includegraphics[width=1.0\linewidth]{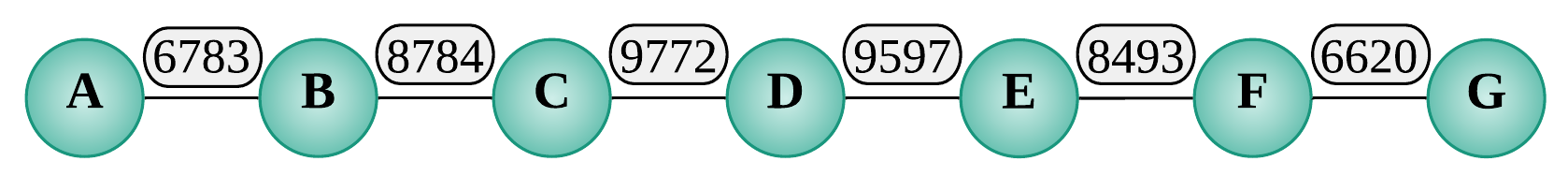}
    \caption{A linear chain of seven spins optimised for both speed and PST. This maps  $\ket{1}_A$ to $\ket{1}_G$ at time $t_f=5.40/J_{max}$ with $99.7\%$ fidelity.}
    \label{fig:linear}
\end{figure}

As these are systems are straightforward and well-understood, in order to test our genetic approach we first check whether or not our algorithm can effectively identify some of the known coupling patterns that enable a spin chain to act as a wire. Our initial configuration is a linear chain of seven spins and we consider the transfer of an excitation being injected at the first or input site (site A in Fig.~\ref{fig:linear}) to its mirror position or output site (here, site G). The quality of the transfer is therefore assessed by calculating the fidelity between the evolved state, $\vert\Psi(t)\rangle$, and the target state of an excitation appearing only at the output site. The transfer is perfect when this fidelity is one.

\begin{figure}[h]
    \centering
    \includegraphics[width=0.9\linewidth]{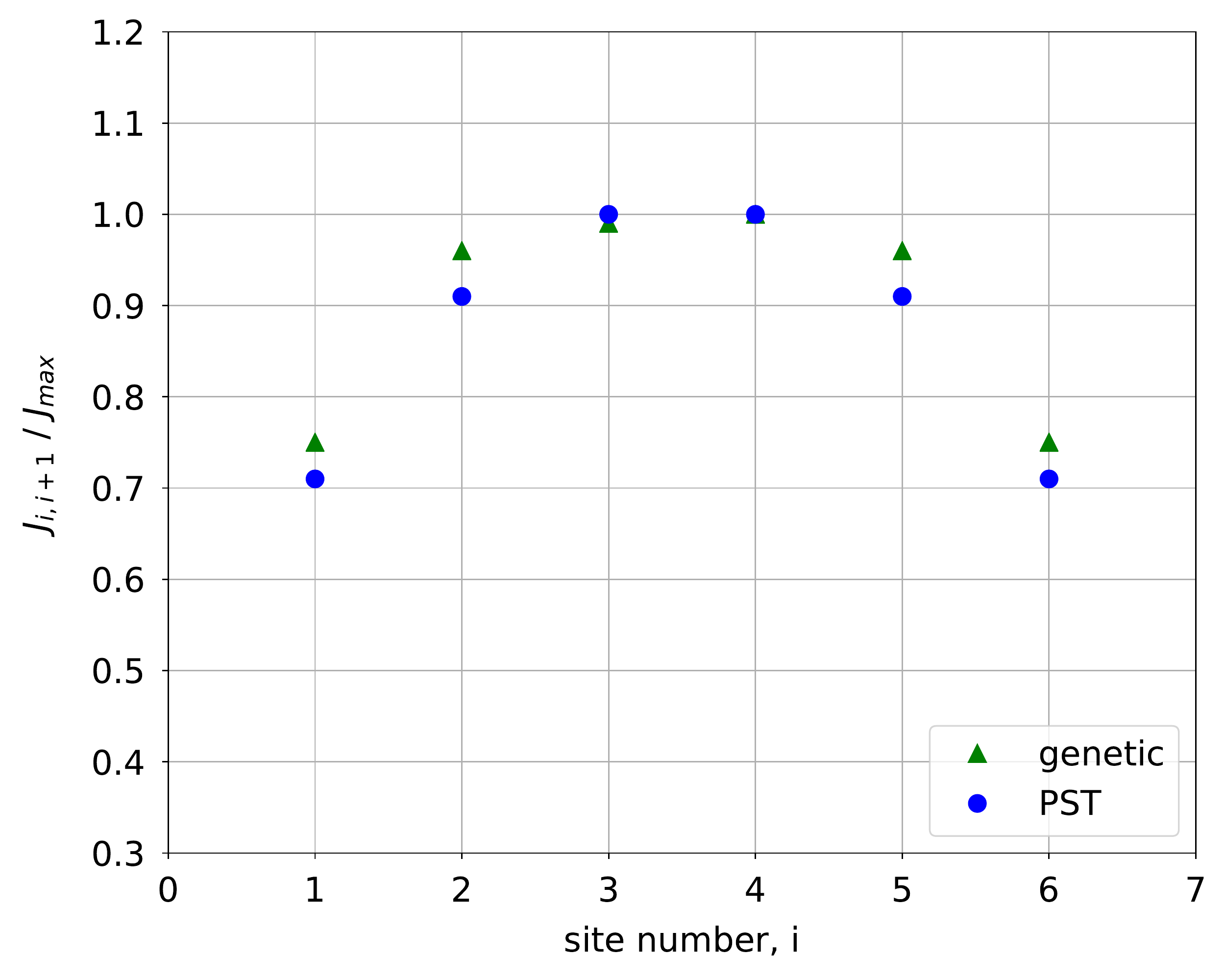}
    \caption{Comparison for a $N=7$ linear chain between the known couplings giving PST from Eq.~\ref{pstcouplings} and the couplings converged upon by the algorithm with default settings. The distribution of generated couplings is flatter than those from Eq.~\ref{pstcouplings} in order to increase speed somewhat at the expense of a small amount of fidelity. Both sets of values are scaled such that their maximum is one.}
    \label{fig:compare}
\end{figure}

The algorithm begins with a generation of genomes all with uniform couplings, then within only a few new generations (see Fig.~\ref{fig:compare}) the genomes converge close to one of the most well-known coupling configurations for PST \cite{christandl2004,niko2004}. Here the $N-1$ interactions of an $N$-site linear chain are defined as,

\begin{equation}
    \label{pstcouplings}
    J_{i,i+1} = J_0 \sqrt{i(N-i)},
\end{equation}

with $i$ being the site number and $J_{0}$ a constant that sets the energy scale. The resultant structure is presented in Fig.~\ref{fig:linear}. We note that as this example utilises only the one-excitation subspace, the last term in the Hamiltonian (Eqn.~1 in the main text) does not contribute, regardless of the value of $\alpha$.

The couplings of the ideal PST chain from Eq.~\ref{pstcouplings} and those obtained from the algorithm are  compared in Fig.~\ref{fig:compare}. Due to the fitness function also having a dependence on speed, the optimized couplings were slightly closer to uniform than those given by Eq.~\ref{pstcouplings}, as shown in Fig.~\ref{fig:compare}, resulting in a slightly faster transfer at the expense of a small amount of fidelity: $t_f=5.4/J_{max}$ with $99.7\%$ fidelity vs $t_f=5.44/J_{max}$ with $100\%$ fidelity.

Changing the parameters of the fitness function would result in a different result being converged upon, allowing flexibility depending on the physical application, e.g.  taking into account factors like the decoherence time and ability to apply error-correction techniques. For instance, in this example, an emphasis on shorter $t_f$ could be included by changing the value of $b$ (from Eq.~3 from the main text): the same optimisation with $b$ set to $-1000$ results in a transfer time of $4.8/J_{max}$, at a fidelity of $92.6\%$.

\subsection{Further Scaling Information}

\begin{figure}[h]
    \centering
    \includegraphics[width=1.0\linewidth]{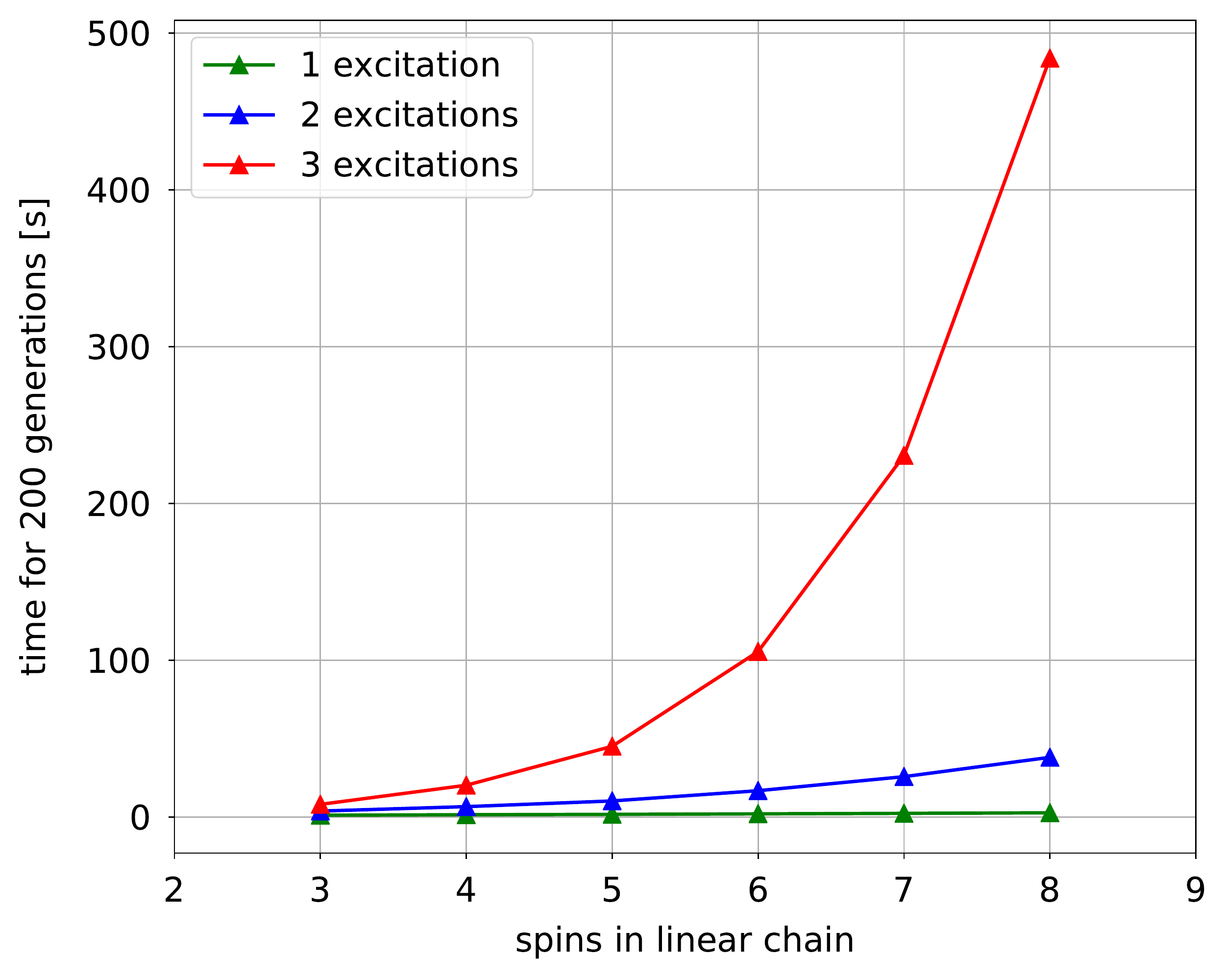}
    \caption{Plot showing how the software scales when increasing the number of sites in the genome for a uniform linear chain as well as increasing excitation subspace size.}
    \label{fig:scalingSitesFixed}
\end{figure}

One important metric regarding scalability is how well the algorithm scales as the number of sites used in the genome increases: this is shown in Figure \ref{fig:scalingSitesFixed}. For this test the algorithm is run for a fixed number of generations in order to eliminate some randomness that would occur if running until a certain fitness score was met. As the number of sites increases, so too does the size of the Hamiltonian matrix and thus each generation takes longer to evaluate the fitnesses due to the increased time spent on diagonalisation. In fact, the dimension of the Hamiltonian is $L\times L$ for the single-excitation subspace ($L$ the number of sites), up to $2^L\times 2^L$ when all excitation subspaces are included in the calculation.  In our examples we are interested in applications that use up to the two-excitation subspace of the entire Hilbert space: two excitations are enough to find a universal set of quantum gates, as this requires two-qubit gates at most. Thus the poorer scaling for three or more excitations is in this case not of major concern.

Given the vast variety of quantum chip topologies that are currently available one could also ask how the efficiency of our algorithm is affected by changing the arrangement of the qubits. To assess this, one would first need to define and fix a task to optimize for. Let us take the example task of single excitation PST over a fixed {\it minimum} number of links between sites A and B (say $m$). Here there is clearly the simplest linear chain solution, and then all manner of more complicated network topologies that may connect the chosen sites, in addition to the most direct $m$-link route. These more complicated networks will all contain an increased number of coupling parameters $k$, with $k>m$, so here it would appear that the ``scaling with topology'' is really set by the scaling with the increased number of couplings in the network. We can conjecture that this scaling relates to that of Fig. 8 of the main text, but now viewed as the scaling for the number of independent couplings in a fixed-task system, rather than the increasing length of a linear chain. The preparation of a separated entangled state, still in just the single excitation subspace, could be expected to scale similarly with the number of couplings, as the topology is changed for this fixed task.

For a different task, such as a gate involving a two-excitation amplitude (all that is required for universality) we would need to specify a minimal circuit that achieves the task and then consider more complicated network topologies also potentially capable of achieving the same task. Here we can conjecture that the ``scaling with topology'' (against the increased number of coupling parameters $k$) will depend on the interplay between the scaling with the two-excitation Hilbert space and the scaling with the number of parameters.

The scaling with the two-excitation Hilbert space is quadratic with the number of network sites $N$ (see Fig. 3 of this Supporting Information). The scaling with the number of parameters $k$ is
exponential (see Fig. 8). For the simplest case of a linear chain topology  $k=N-1$; however  $k=N(N-1)/2$ for a fully connected graph topology.  The interplay will thus depend on the relation between $N$ and $k$ in the various topologies considered.

\subsection{Visualising Optimisation}

\begin{figure}[h]
    \centering
    \includegraphics[width=1.0\linewidth]{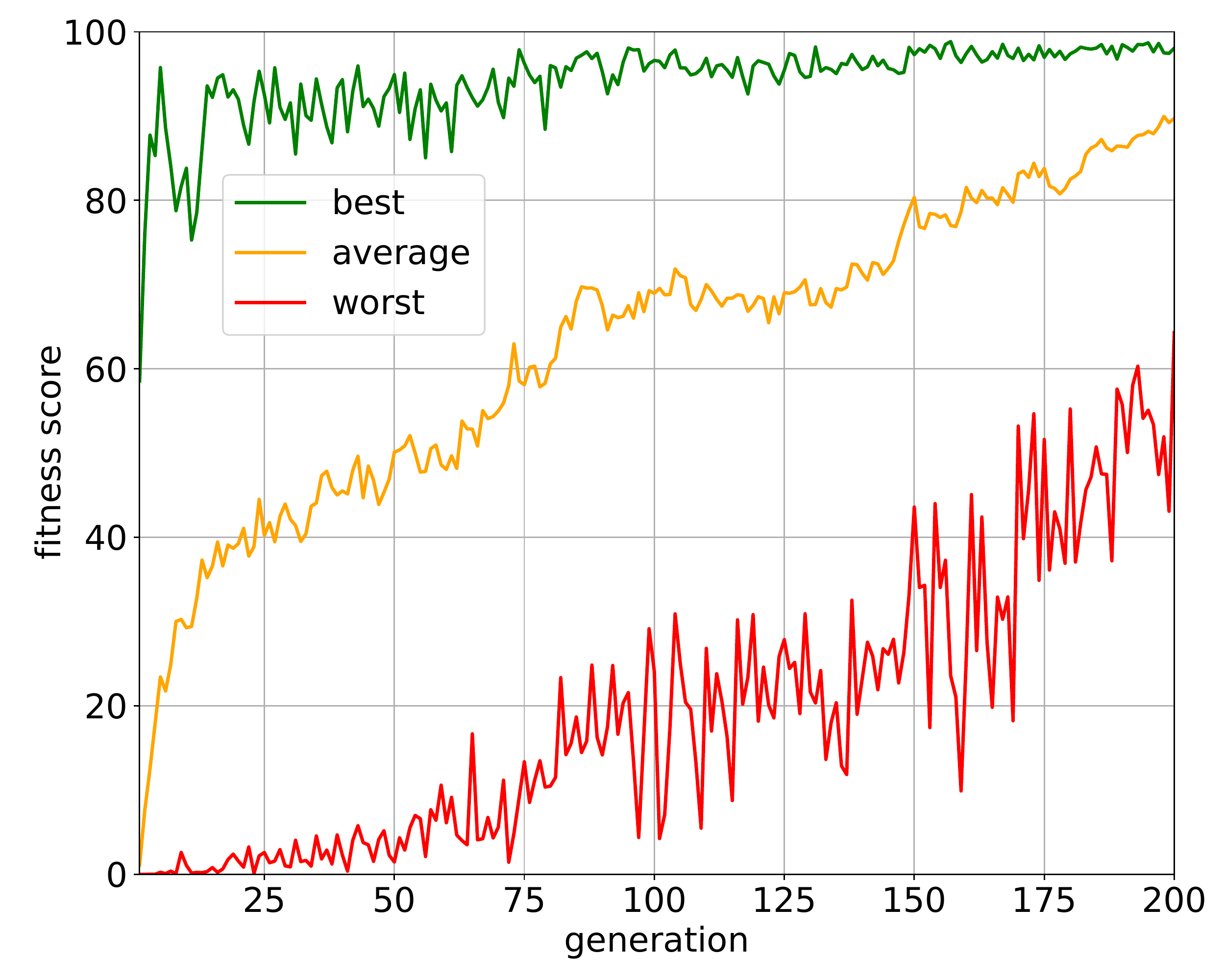}
    \caption{Plot of the worst, average and best fitness scores $f(F_\text{max},t_f)$ for each generation when optimising a seven-site linear chain for PST. The size of mutations is reduced each generation, resulting in smaller, more precise, changes in fitness. This is the case except for the worst case scenario, where sometimes even small changes in the couplings results in large changes to fitness, which represents how the parameter space is being explored.}
    \label{fig:geneticChain}
\end{figure}

\begin{figure}[h]
    \centering
    \includegraphics[width=1.0\linewidth]{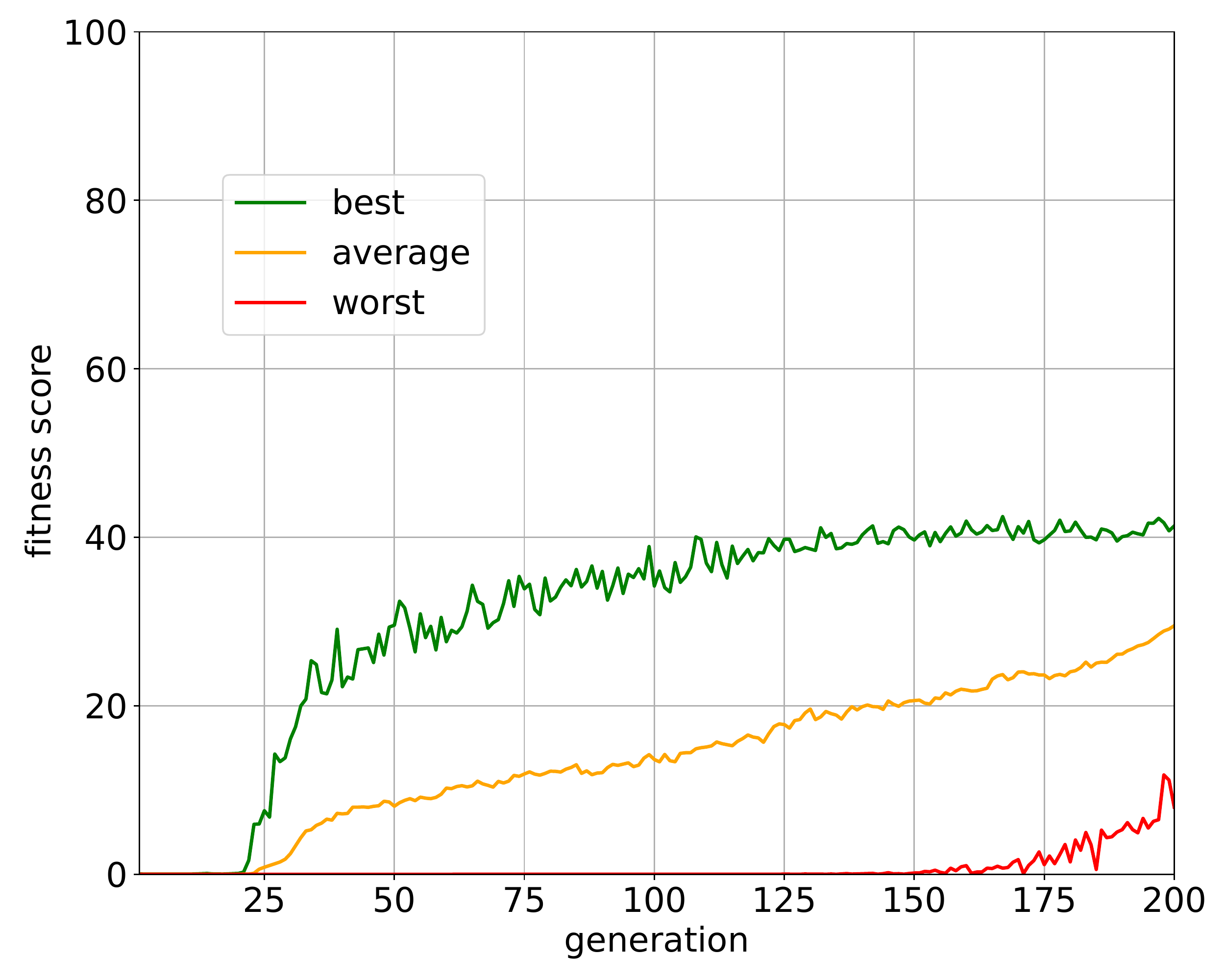}
    \caption{Plot of the worst, average and best fitness scores $f(F_\text{max},t_f)$ for each generation when optimising a $4\times 4$ grid of qubits to perform a controlled phase gate. The size of mutations is reduced each generation, resulting in smaller, more precise, changes in fitness, except for the worst fitness scores, which are the product of more diverse genomes. The final fitness score here corresponds to around $91\%$ fidelity: further optimisations with reduced maximum mutation sizes were then performed to reach the result of $99.8\%$ given in the main text.}
    \label{fig:geneticPhase}
\end{figure}

During optimisation our program maintains a log of the worst, average and best fitness scores of each generation. Graphs of such indicators from the optimisations of the PST linear spin chain and phase-gate examples from the main text are given in Fig.~\ref{fig:geneticChain} and Fig.~\ref{fig:geneticPhase}, which shows how few generations are required for this method to reach a high fitness score for such simple systems. Note that here the optimisation was stopped after the default maximum number of generations (200) as a demonstration, but could have been stopped much sooner and still produced fast and high fidelity transfer.

\begin{figure}[p!]
    \centering
    \begin{subfigure}{1.00\linewidth}
        \includegraphics[width=1.0\linewidth]{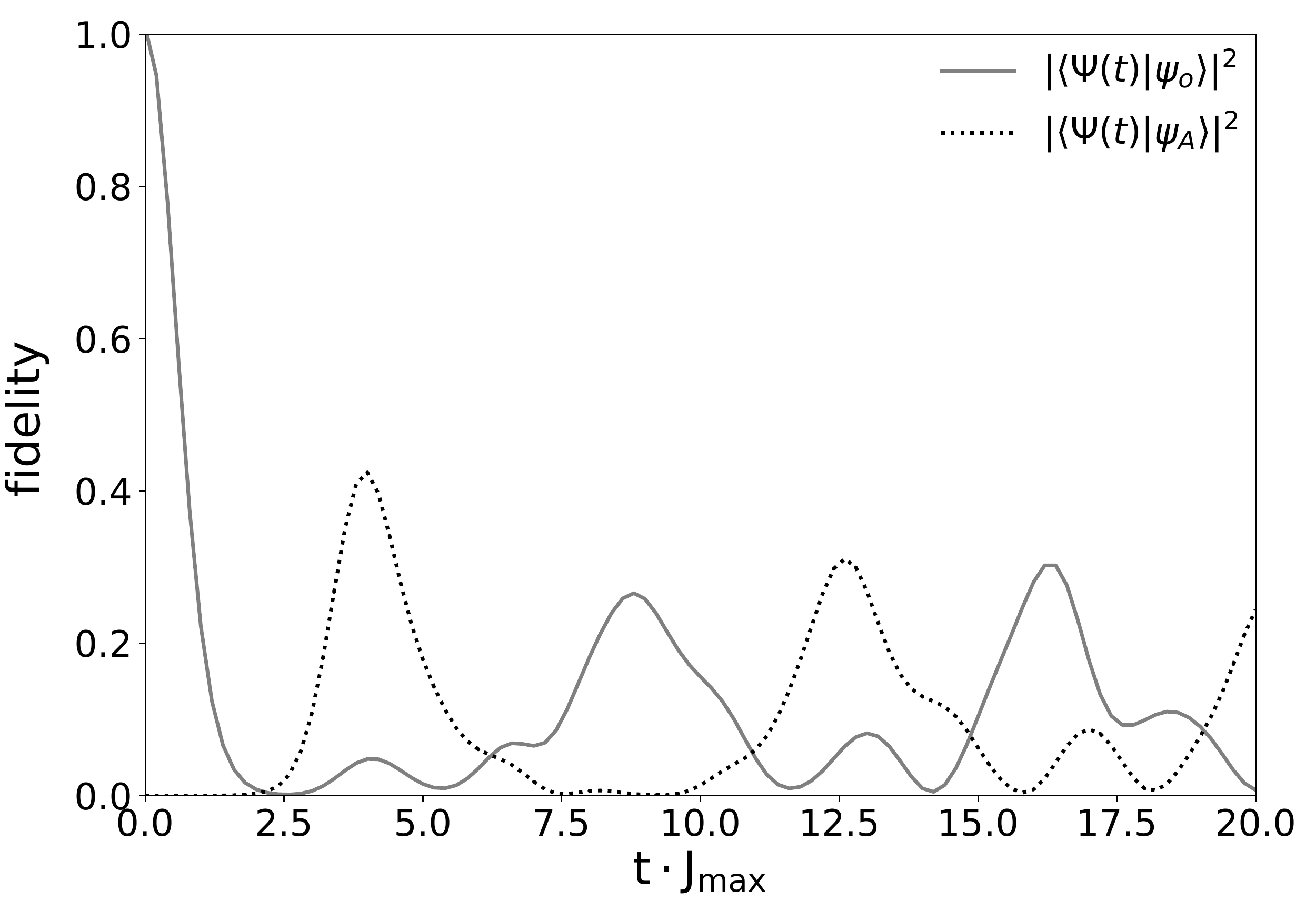}
    \end{subfigure}
    \begin{subfigure}{1.00\linewidth}
        \includegraphics[width=1.0\linewidth]{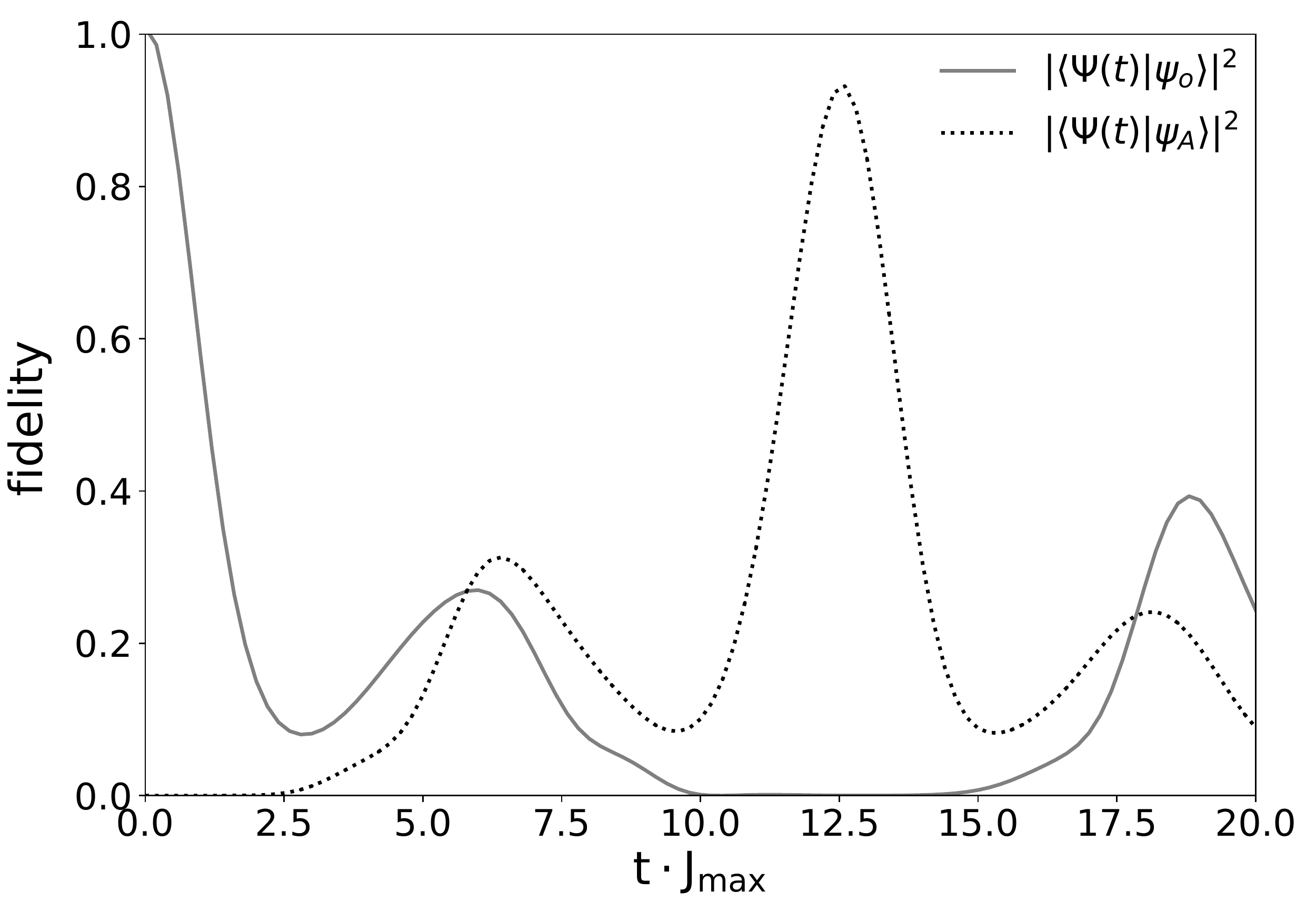}
    \end{subfigure}
    \caption{Plots of fidelity $F(t)$ vs time, both before (top, uniform couplings) and after (bottom) the optimisation of the phase gate. The fidelity with respect to the initial state is the solid line, whilst the fidelity with respect to the desired final state is the dashed line.}
    \label{fig:beforeafter}
\end{figure}

 For more complex topologies like the $4\times 4$ grid used to create the phase gate, the dynamics of the spin network corresponding to the input state for the optimization process (uniform couplings)  may show very unpredictable behaviour with low fidelities. After some time spent optimising, however, the graph of the fidelity against time becomes more regular, as a high fidelity peak begins to emerge within the desired time window. This is shown in figure \ref{fig:beforeafter}, which compares the fidelities before (with uniform couplings) and after optimisation of the network for the phase gate system.

\subsection{Genome Visualisation}

\begin{figure}[t]
    \centering
    \includegraphics[width=0.6\linewidth]{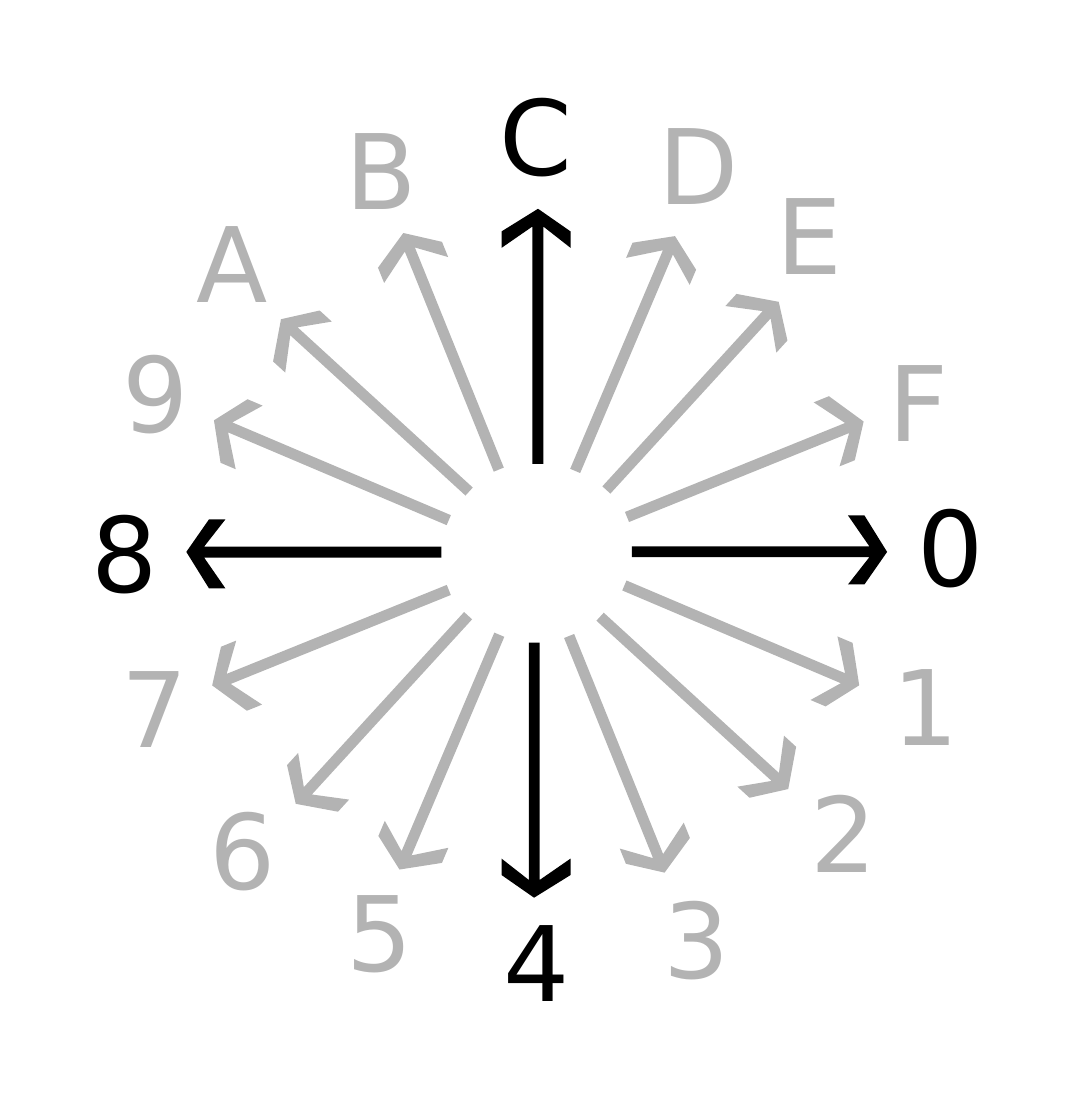}
    \caption{Diagram showing how the cardinal directions map to hexadecimal characters in the genome visualisation information.}
    \label{fig:wheel}
\end{figure}

At the end of the string representing the genome we include optional information only relevant when creating a visualisation of the network. This is given as a series of hexadecimal characters after a hash sign. Each of these characters represents the angle that a corresponding coupling (in the same order as in the main bulk of the genome) should be placed at with respect to a given site, for example a $\#0$ represents a rightwards coupling, whilst $\#4$ is downwards. This is shown with more detail in figure \ref{fig:wheel}. The difference between each of the integers in the circumference are represented by $22.5^{\circ}$ increments.

In order to reconstruct the genome visually, the program loops over the couplings in the main section of the genome and attempts to place any coupling which has not yet been placed. This coupling is then placed at the angle given by the corresponding visual info character at the end of the genome. This method works well as a simple and compact way of allowing such complex genomes to be viewed in a human-readable manner.

\end{document}